\documentclass[fleqn,usenatbib,useAMS]{mnras}

\usepackage{graphicx}	
\usepackage{amsmath}	
\usepackage{amssymb}	
\usepackage{multicol}        
\usepackage{bm}		
\usepackage[compatibility=false]{caption}
\usepackage{subcaption}

\usepackage[T1]{fontenc}
\usepackage{ae,aecompl}

\usepackage{newtxtext,newtxmath}

\title[Dust evolution in photoevaporating discs] {The evolution of dust in discs influenced by external photoevaporation}

\author[A. D. Sellek et al.]{
Andrew D. Sellek$^{1}$\thanks{Contact e-mail: \href{mailto:ads79@cam.ac.uk}{ads79@cam.ac.uk}}
Richard A. Booth$^{1}$
Cathie J. Clarke$^{1}$
\\
$^{1}$Institute of Astronomy, University of Cambridge, Madingley Road, Cambridge CB3 0HA, UK}

\date{Last updated 2019 December 11; in original form 2019 October 18}

\pubyear{2019}

\begin{document}
\label{firstpage}
\pagerange{\pageref{firstpage}--\pageref{lastpage}}
\maketitle

\begin{abstract}
Protoplanetary discs form and evolve in a wide variety of stellar environments and are accordingly exposed to a wide range of ambient far ultraviolet (FUV) field strengths. Strong FUV fields are known to drive vigorous gaseous flows from the outer disc. In this paper we conduct the first systematic exploration of the evolution of the solid component of discs subject to external photoevaporation. We find that the main effect of photoevaporation is to reduce the reservoir of dust at large radii and this leads to more efficient subsequent depletion of the disc dust due to radial drift. Efficient radial drift means that photoevaporation causes no significant increase of the dust to gas ratio in the disc. We show that the disc lifetime in both dust and gas is strongly dependent on the level of the FUV background and that the relationship between these two lifetimes just depends on the Shakura-Sunyaev $\alpha$ parameter, with the similar lifetimes observed for  gas and  dust in discs pointing to higher $\alpha$ values ($\sim 10^{-2}$). On the other hand the distribution of observed discs in the plane of disc size versus flux at $850~\mu$m is better reproduced by lower $\alpha$ ($\sim 10^{-3}$). We find that photoevaporation does not assist rocky planet formation but need not inhibit mechanisms (such as pebble accretion at the water snow line) which can be effective sufficiently early in the disc's lifetime (i.e. well within a Myr).

\end{abstract}

\begin{keywords}
protoplanetary discs -- planets and satellites: formation -- submillimetre: planetary systems
\end{keywords}



\section{Introduction}
It is apparent from a number of lines of evidence that protoplanetary discs are subject to environmental damage from far ultraviolet (FUV) radiation.
This is seen most readily in regions of very high FUV background, such as the Orion Nebula Cluster, where proplyd structures, consistent with ionisation by radiation from  the cluster's most luminous O star, are observed in the vicinity of young low mass stars \citep{Odell_1993,Henney_1998}.  The few hundred au offset of these structures from their parent stars  provides  evidence for strong neutral disc winds, driven by environmental FUV radiation, which prevent the penetration of ionising photons to the disc surface \citep{Johnstone_1998}, an  interpretation  corroborated by spectroscopic measurements \citep{Henney_1999,Henney_2002}.
{\it Indirect} evidence for the role of FUV irradiation in depleting protoplanetary discs is provided by the observed inverse correlation between FUV flux and the incidence of protoplanetary discs observed in the Cygnus OB2 association \citep{Guarcello}.

While the fraction of young stars in the Galaxy that are subject to the high ultraviolet fluxes experienced in OB associations and dense clusters is only a few tens of per cent \citep{Fatuzzo_2008}, there is now increasing interest in the possibility of significant environmental damage also in regions of relatively low ultraviolet background
{\footnote{The FUV background is conventionally denoted as a multiple of the Habing unit, ${\rm G_0}$, which is $1.6 \times 10^{-3}$ erg cm$^{-2}$ over the energy range $6-13.6$ eV \citep{Habing_1968}, such that the local interstellar field is $1.7~{\rm G_0}$}}.
Following \citet{Adams_2004}, \citet{Facchini_16} constructed models for FUV driven winds down to much lower FUV fields, showing that even here  winds provide an important sink of disc mass for discs larger than $\sim 100$ au. Subsequently, \citet{Haworth_2017} modeled the extended CO halo around the large disc in IM Lupi in terms of a photeoevaporative wind driven by the mild ($\sim 4~{\rm G_0}$) FUV background in its vicinity \citep{Cleeves_2016}.

While a number of studies have modeled the impact of photoevaporation on protoplanetary disc demographics \citep{Scally,Winter_2018,Winter_2019, Concha_Ramirez,Nicholson} such modeling has studied the dispersal of disc {\it gas} and then compared with observational diagnostics which are, however, mainly based on disc {\it dust}.
Such an approach is reasonable if disc dust grains remain small and thus well coupled to the gas. However grain growth causes partial dynamical decoupling between dust and gas. This both a) limits the capacity of the wind to remove dust \citep{Throop,Facchini_16,Hutchison_2016,Carrera_2017} but also b) leads to radial drift of dust that is torqued down by the mildly sub-Keplerian rotation of the disc gas \citep{Whipple_1973,Weidenschilling_1977,Takeuchi_2005,Brauer_2008,Birnstiel_2010,Birnstiel_12}.
While the former enhances the dust to gas ratio, and has been invoked as a possible instigator of conditions necessary for triggering the streaming instability \citep{Youdin_2005} in the residual dust layer, the latter instead lowers the dust to gas ratio \citep{Birnstiel_12} unless some mechanism (such as pressure traps introduced by planets; \citealt{Paadekooper_2004, Rosotti_2016}) acts to retain dust grains in the outer disc \citep{Ricci_2012,Pinilla_2012}. To date the only calculation which has attempted to disentangle these two trends in the case of external, FUV driven photoevaporation is that of \citet{Haworth_18_Trappist} which supplemented a calculation of the gas evolution with  estimates of the growth and migration of the  dust component.

In the present paper we remedy this insufficiency by conducting calculations of viscously evolving protoplanetary discs that are subject to photoevaporation and where we model the evolution of the dust component using the methodology of \citet{Birnstiel_12} (see also \citealt{Booth_17}). This approach models dust growth and its limitation by radial drift or fragmentation and tracks the evolution of the dust surface density as a result of radial drift and partial entrainment in the photoevaporative wind. Our chief motivations for this study are: i) to compare the lifetimes of the gas and dust in protoplanetary discs as a function of radiative environment ii) to determine how photoevaporation affects the sizes of dust discs in order to compare with high resolution submm surveys in a variety of environments \citep[c.f.][]{Tazzari_17,Barenfeld,Tripathi} and iii) to consider how photoevaporation affects the planet formation potential of protoplanetary discs. In the latter regard, we will focus on the effect on the solid component, assessing whether the assembly of rocky planets is promoted by enhancement of the local dust to gas ratio in the outer disc \citep{Throop} or whether (as argued by \citealt{Haworth_18_Trappist}) the loss of dust in the wind at early times instead  suppresses rocky planet formation.
 
In this study we take advantage of the recently published FRIED grid of photoevaporation models \citep{Haworth_18_FRIED} which presents photoevaporation rates as a function of disc outer radius, local density and stellar mass over a wide range of background field strengths. These 1D thermochemical simulations (in which mass loss is concentrated at the disc's outer rim) have been calibrated against 1D  analytic solutions \citep{Haworth_2016} and are also found to be in broad agreement with the results of 2D simulations \citep{Haworth_2019}, although the latter yield  somewhat larger total mass loss because of mass loss from the disc's upper and lower surfaces.  We detail the coupling of the photoevaporation grid to the viscous evolution of the disc and the evolution of the dust component in Section 2. In Section 3 we briefly describe the evolution of the disc gas under combined viscous evolution and photoevaporation and demonstrate the same qualitative features (using a previous photoevaporation prescription) found by \citet{Clarke_2007}. In Section 4 we describe a fiducial model containing dust, demonstrating an early phase of dust growth and entrainment in the wind followed by a successive phase of dust depletion due to radial drift. In Section 5 we explore the dependence of the dust evolution on model parameters and in Section 6 we summarise the implications of our modeling for reproducing the observed demographics of protoplanetary discs. Section 7 discusses how our results affect the planet forming potential of protoplanetary discs and Section 8 summarises our conclusions. 
 
\section{Model}

\subsection{Treatment of viscous  evolution and dust evolution }
We model the viscous evolution of the gas and growth and radial migration of the dust as in \citet{Booth_17} for which the treatment of the dust follows that of \citet{Birnstiel_12}. We employ a mid-plane radial temperature profile $T(R)$ modeled
according to:
\begin{equation}
    T = T_0 \left( \frac{R}{R_0} \right)^{-q}
 \label{eq:T_law}   .
\end{equation}
Assuming a constant $\alpha$ viscosity prescription \citep{Shakura-Sunyaev}, the kinematic viscosity $\nu$ then scales with radius as
\begin{equation}
    \nu \propto R^{\gamma_{\rm visc}} \propto R^{3/2-q}
    \label{eq:visc_law}
    .
\end{equation}
We explore spatially uniform  $\alpha$ values motivated by the range of viscosities inferred from observations: $10^{-3}$ \citep{Rafikov_2017} and $10^{-2}$ \citep{Hartmann_1998}.
Following \citet{Clarke_2007} and \citet{Facchini_16} we then choose $q=0.5$ such that $\nu$ is linear with radius. This dependence - which results if stellar irradiation of dust in the form of the blackbody equilibrium temperature dominates the heating \citep{Kenyon_Hartmann_1987,Chiang_Goldreich_1997} - is supported by mean fits of the SED shape for $q$ \citep{Andrews_2005,Kenyon_Hartmann_1995} and the corresponding linear dependence of viscosity on disc radius is also consistent with the observed variation of accretion rates onto T Tauri stars as a function of age \citep{Hartmann_1998}.

In order to set a temperature scale, we set the aspect ratio (the ratio of scale height $H$ over radius, which scales as $H/R \propto R^{1/4}$) to $H/R \sim 0.033$ at $R_0 = 1~\mathrm{au}$ such that $T_0\approx279~\mathrm{K}$ for a $1~{\rm M_{\sun}}$ star.
As our initial condition we adopt the similarity solution of
\citet{LBP_1974} for the surface density, $\Sigma$:

\begin{equation}
    \Sigma = \Sigma_0 \left(\frac{R}{R_{\rm C}}\right)^{-\gamma} \exp\left( -\frac{R}{R_{\rm C}} \right)^{2-\gamma}
    \label{eq:LBP_profile}
    ,
\end{equation}
where $\gamma = \gamma_{\rm visc}$ (Equation \ref{eq:visc_law})
and $R_{\rm C}$ is the scale radius of the exponential cut off (i.e. $R_{\rm C}$ sets the initial extent of the disc).
$R_{\rm C}$ was varied between $10$ and $300$ au. The values adopted are roughly in the range derived from submm continuum observations of protoplanetary discs \citep[e.g.][]{Ansdell_2018,Tripathi}, although this observational measure should not necessarily coincide with the initial, mass-based measure of radius represented by $R_{\rm C}$.
The density normalisation, $\Sigma_0$, is set by the total disc mass. Observational studies measure \textit{dust} masses of up to $\sim 100~M_\oplus$ \citep[e.g. $141~M_\oplus$ in Lupus,][]{ALMA_Lupus}, corresponding to discs with $\gtrsim 30$ Jupiter masses ($M_{\rm J}$) total mass, assuming the canonical gas-to-dust ratio of $100$. Dust masses are known to decline with age \citep{ALMA_Orionis} so we start our discs with a range of masses $\leq 100~M_{\rm J}$, this being close to the maximum permitted value for gravitational stability.

For $q=0.5$, $\gamma=1$ and thus the surface density  scales as $\Sigma \propto R^{-1}$ for $R<R_{\rm C}$.
As part of our parameter exploration in Section \ref{sec:parameters}, we relax the assumptions that $q=0.5$ and that the disc starts in the similarity solution appropriate to its viscosity (i.e. with $\gamma = \frac{3}{2} - q$), instead letting it adjust to the steady state.
In all cases, we evolve the viscous diffusion equation for the gas using a grid equispaced in $R^{1/2}$ between $R_{\mathrm{in}} = 0.1~\mathrm{au}$ and $R_{\mathrm{out}} = 400~\mathrm{au}$.

We follow the \citet{Birnstiel_12} model in approximating the dust as two populations:
a \textit{small population} of fixed monomer size $a_0 =0.1~\mathrm{\mu m}$ at which all the dust starts, 
and
a \textit{large population} whose maximum size $a_{\mathrm{max}}$ is governed by prescriptions for particle growth and limits  set by either fragmentation or radial drift: we adopt a fragmentation velocity $u_{\rm f}=10$ m s$^{-1}$ \citep{Gundlach_2015} and an internal dust density of  $\rho_{\rm s} = 1~\mathrm{g~cm}^{-3} $ throughout.
The bulk density depends on the composition and porosity of the grains and here we adopt a value appropriate to the composition used in \citet{Tazzari_2016}
\footnote{By volume: 5.4\% astronomical silicates,  20.6\%  carbonaceous  material,  44\%  water ice, and 30\% vacuum, which gives an average dust grain density of $\sim 1~\rm g~cm^{-3}$.}.
The initial mass fraction of dust is taken to be the canonical value of 1\%, inherited from the interstellar medium \citep{Bohlin_1978}.
 
Following the evolution of the maximum grain size in the large population, the code re-normalises the grain size distribution in the large population so as to maintain a fixed mass proportion in each population. 
Each dust population is subject to advection with the  viscous flow of the gas plus radial drift with respect to the gas and diffusion which are prescribed as a function of Stokes number $St$ (ratio of drag time to orbital time).  In the Epstein drag regime, $St$  is given, for a grain of size $a$,  by:
 \begin{equation}
    St = \frac{\pi}{2} \frac{a \rho_{\rm s}}{\Sigma}
    \label{eq:Stokes_def}
    .
\end{equation}
 (see \citet{Birnstiel_12} for further details). We follow  
 \citet{Booth_17} and  \citet{Tanaka_2005} in incorporating additional terms in the
 gas evolutionary equations that account for the back-reaction of the dust on the gas although, given the consistently low dust to gas ratios in our models, this is never of any practical importance to the system evolution.

\subsection{Treatment of photoevaporation}
\subsubsection{Calculating the Mass Loss Rates}
 
\begin{figure}
    \centering
    \includegraphics[width=\linewidth]{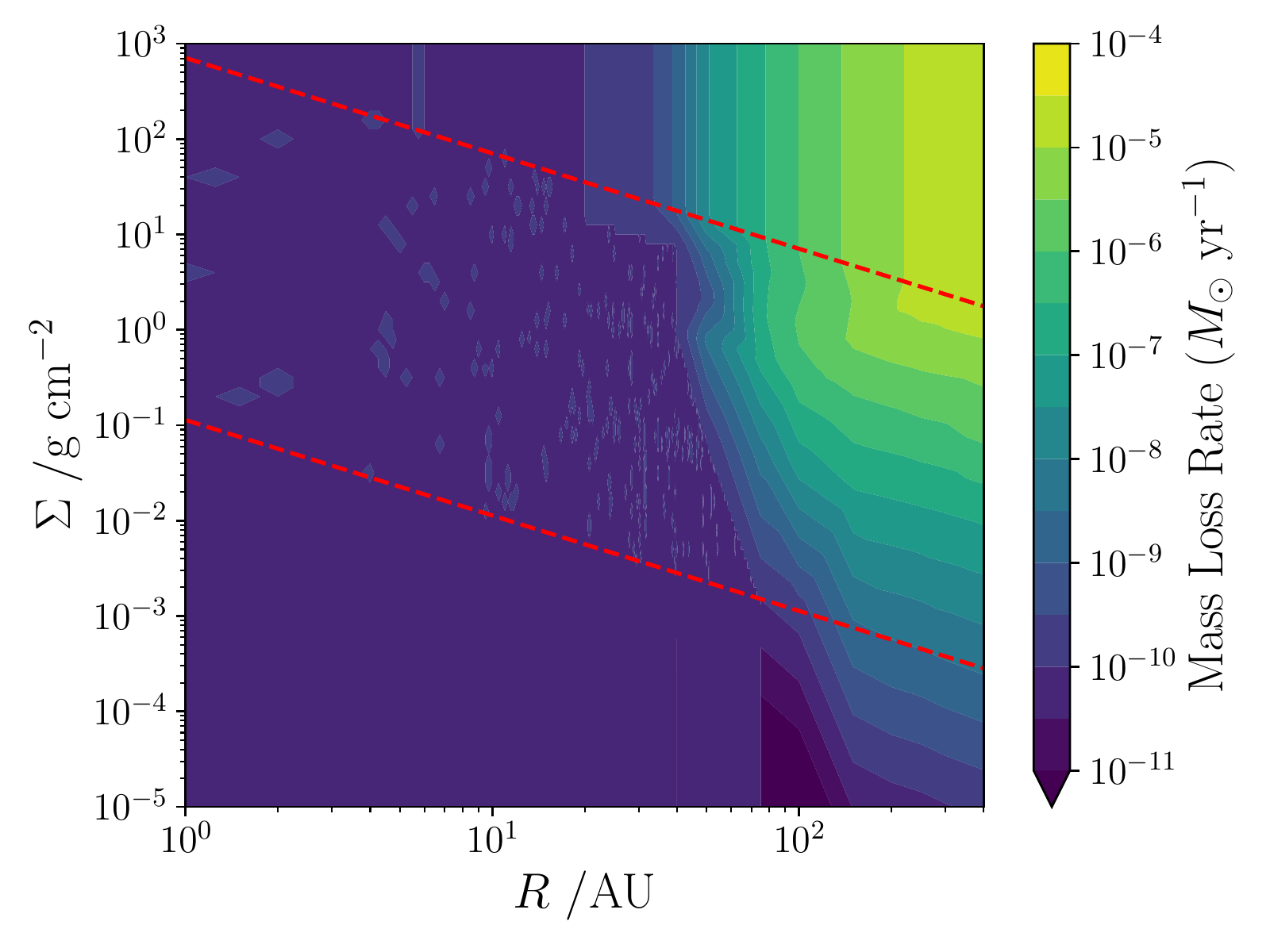}
    \caption{Interpolation and extrapolation of the mass loss rates.
    To ensure a smooth variation, the interpolation was performed on $R_D$ and the transformed variable $M_{400} = 2\pi \Sigma_{\rm out} R_{\rm D}^2 \left(\frac{R_{\rm D}}{400~\rm au}\right)^{-1}$ for a star with $M_* = 1~\rm{M_{\sun}}$ and an FUV flux of $1000~{\rm G_0}$. High mass loss rates (yellow colours) are seen at large radii and high surface densities while lower mass loss rates (darker colours) are seen at small radii or lower surface densities. The red dashed lines represent the limits of the original \citet{Haworth_18_FRIED} FRIED grid, outside of which we extrapolate the values according to Equation \ref{mdotextrap}.}
    \label{fig:interpolation}
\end{figure}

We used the scipy LinearNDInterpolator \citep{scipy}  in order to extract photoevaporative mass loss rates (as a function of outer disc radius, $R_{\rm D}$, outer  disc surface density, $\Sigma_{\rm out}$, ultraviolet flux and stellar mass) from the FRIED grid of \citet{Haworth_18_FRIED}.
Note that the grid has an artificially imposed lower limit on the mass loss rates of $\dot{M} = 10^{-10}~{\rm M_{\sun} yr}^{-1}$ because the simulations are unreliable below this threshold. At a given outer disc radius, these rates are  insensitive to $\Sigma_{\rm out}$ in the limit of high $\Sigma_{\rm out}$ where the wind is optically thick to the incoming FUV radiation but vary linearly with $\Sigma_{\rm out}$ at low $\Sigma_{\rm out}$ where  the wind is instead optically thin to the ambient FUV field \citep{Facchini_16}.
For this reason we adopt the following interpolation scheme at surface densities outside the range between the lower and upper limits - $\Sigma_{\mathrm{min}}(R)$ and $\Sigma_{\mathrm{max}}(R)$  (both of which scale as  $R^{-1}$) - of the FRIED grid:
 
\begin{equation}
    \dot{M}(R,\Sigma) =
    \begin{cases}
        \dot{M}_{\mathrm{max}}(R)
        & \Sigma > \Sigma_{\mathrm{max}} \\
        \dot{M}_{\mathrm{min}}(R) \frac{\Sigma}{\Sigma_{\mathrm{min}}}
        & \Sigma < \Sigma_{\mathrm{min}}, \dot{M}_{\mathrm{min}}>10^{-10} M_{\sun} \mathrm{yr}^{-1} \\
        10^{-10} M_{\sun} \mathrm{yr}^{-1}
        & \Sigma < \Sigma_{\mathrm{min}}, \dot{M}_{\mathrm{min}}=10^{-10} M_{\sun} \mathrm{yr}^{-1}
    \end{cases}
    \label{mdotextrap}
\end{equation}
where $\dot{M}_{\mathrm{min}}(R) = \dot{M}(R, \Sigma_{\mathrm{min}}(R))$ and $\dot{M}_{\mathrm{max}}(R) = \dot{M}(R, \Sigma_{\mathrm{max}}(R))$. Fig. \ref{fig:interpolation} presents an example grid for a solar mass star with FUV flux of $1000~{\rm G_0}$ where the red lines denote the limits of parameter space calculated by \citet{Haworth_18_FRIED} and the rest of the grid is generated by the above extrapolation.

\subsubsection{Implementation}
A numerical issue in modeling external photoevaporation is that it is necessary to define the outer edge of the disc in order to prescribe the appropriate mass loss rate. If the results of the FRIED grid are applied at face value then, given a typical surface density profile in the disc, the photoevaporation rate would typically increase with chosen outer disc radius as long as the wind is optically thick in the FUV (because the photoevaporation rate increases in weakly bound outer regions) and would then decrease for larger chosen disc radii in the optically thin regime (where the decline is driven by the linear scaling of wind mass loss rate with surface density). This behaviour would imply that the importance of photoevaporation could be very sensitive to the numerical implementation (i.e. to the surface density threshold defining the disc outer edge).
 
\begin{figure}
    \centering
    \includegraphics[width=\linewidth]{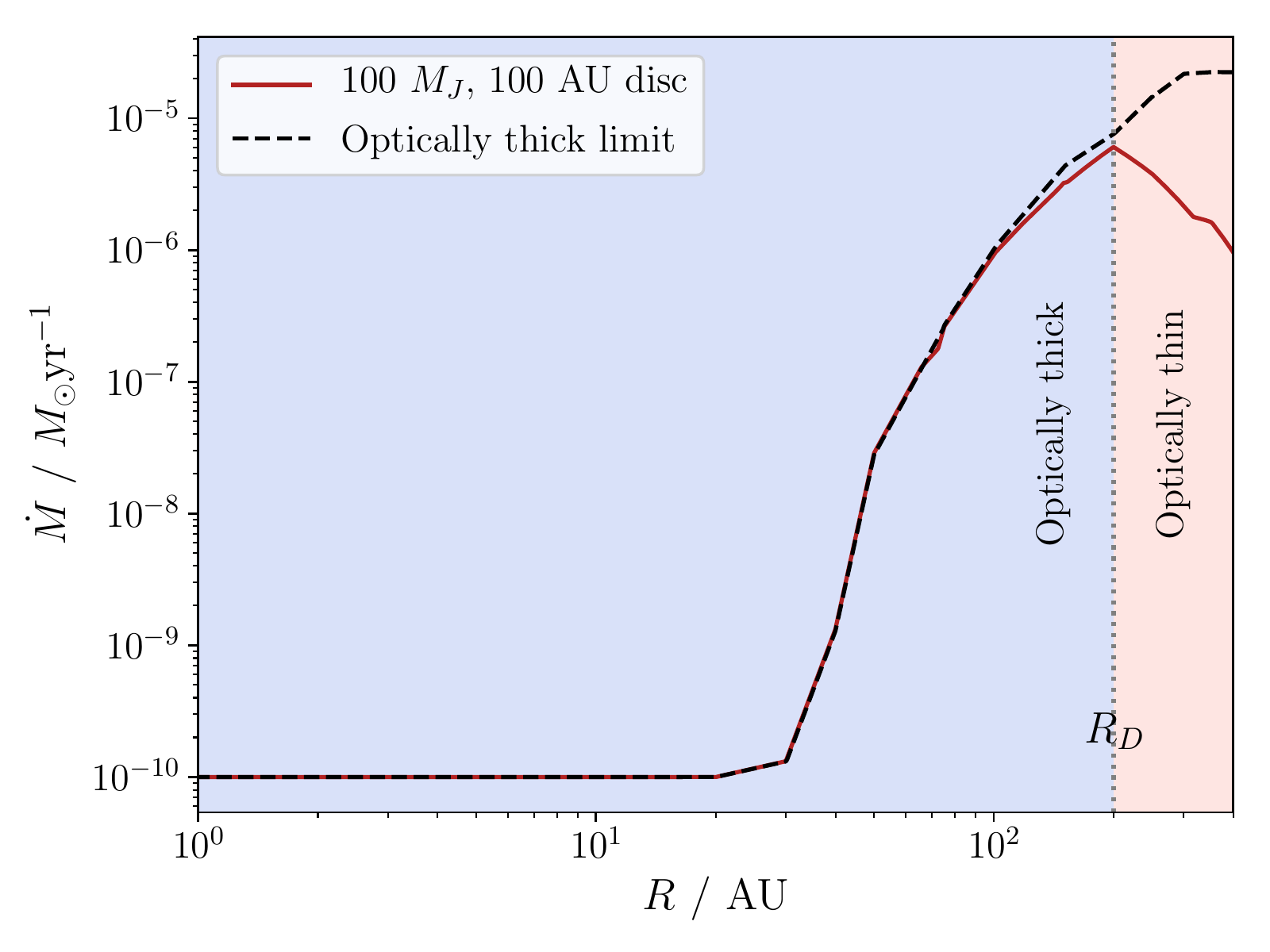}
    \caption{ Mass loss rates for $1000~G_0$ from the FRIED grid (red solid curve) as a function of designated outer disc radius for a disc with surface density profile given by equation \ref{eq:LBP_profile}, scale radius $R_{\rm C}=100$ au and total gas mass $M_{\rm D} = 100$ Jovian masses. The dashed line gives the corresponding FRIED rates if the wind were optically thick at all radii, demonstrating that the peak in the mass loss rate corresponds to the radius where the wind optical depth is of order unity. In our prescription, the outer edge of the disc, $R_{\rm D}$ and corresponding mass loss rate is set to the value corresponding to this maximum.}
    \label{fig:dotM_R_example}
\end{figure}

However this uncomfortable conclusion neglects the fact that the optically thin FUV rates are not physically self-consistent because they are calculated assuming that the material within the disc outer radius is not allowed to evolve in response to external FUV heating. This assumption  is not sustainable if the wind is optically thin to the FUV since in reality the flow would be set by the larger flow rates from smaller radius. In practice this means that the effective outer radius of the disc, which sets the mass loss rate, is located at the optically thick/thin transition, i.e. where the nominal mass loss rate from the FRIED grid attains a maximum value.
We illustrate  (red line in  Fig. \ref{fig:dotM_R_example}) the dependence of the nominal (FRIED) photoevaporation rate as a function of designated outer radius for the initial surface density profile of a disc with $R_{\rm C}=100$ au and disc mass of $100~\rm{M_{J}}$, and contrast the decline, beyond $\sim 200$ au, with the mass loss in the limit of an optically thick FUV wind (black dashed). In this example, the effective outer radius and corresponding mass loss rate would, therefore, in our prescription, be set to $\sim 200$ au.
  
Hence, at every timestep we evaluate  the \textit{disc radius}, $R_{\rm D}$, from the instantaneous surface density profile as the radius where the mass loss rate $\dot{M}(R)$ (the `FRIED' mass loss rate calculated at each grid point as though this were the outer radius of the disc) is maximal.
 
Having located the region of the disc from which the mass loss is originating we calculate the `total mass loss'  rate via a mass weighted prescription over cells exterior to $R_{\rm D}$ ($M_i$, $\dot{M}_i$ are the mass and mass loss rate of the cell i):
\begin{equation}
    \dot M_{\mathrm{tot}} = \sum_i \dot{M}_i \frac{M_i}{M(R>R_{\rm D})}
    \label{eq:mdot_tot}
    .
\end{equation}
The purpose of this step is to avoid large numerical fluctuations in the mass loss rate as cells are emptied: in practice $\dot M_{\mathrm{tot}}$ will be dominated by the cell at $R_{\rm D}$ (i.e. the maximum mass loss rate), as this has both the largest mass and largest mass loss rate but as this cell is emptied adjacent cells start to contribute to the mass loss rate.

Having evaluated the mass loss rate we need to decide the range of cells to which this is applied. 
To ensure that we can't leave mass outside of the wind base indefinitely, we apply this mass loss rate spread across all cells with $R>R_{\rm D}$ (i.e. outside of the disc's designated radius) in proportion to the mass they contain.
\begin{equation}
\dot M_{\mathrm{eff},i} = \dot M_{\mathrm{tot}} \frac{M_i}{M(R>R_{\rm D})}
    .
\end{equation}

Note that the models were stopped when no mass loss rate was found in the grid that was above the floor  of $10^{-10}~M_{\sun}~\mathrm{yr}^{-1}$ imposed by \citet{Haworth_18_FRIED}.

\subsubsection{Photoevaporation of Dust}
 Photoevaporation results in  a wind whose radial velocity $v_r$  relates to the mass loss rate $\dot{M}$ through \citep{Adams_2004}
\begin{equation}
     \dot{M} = 4 \pi R^2 \mathcal{F} \rho_{\rm g} v_r
     .
\end{equation}
This represents a spherical radial flow over a solid angle $4 \pi \mathcal{F}$, where the geometric factor $\mathcal{F} = \frac{H}{\sqrt{R^2+H^2}} $ tends towards the aspect ratio ($H/R$) for thin discs.
This radial motion produces a drag force on a dust particle of radius $a$ of
\begin{equation}
    F_{\rm D} = \frac{1}{3} a^2 v_{\rm th} \frac{\dot{M}}{R^2 \mathcal{F}}
    ,
\end{equation}
where $v_{\rm th}$ is the thermal velocity of the gas.
For a star of mass $M_*$, the gravitational force on a dust grain is
\begin{equation}
    F_{\rm G} = - \frac{GM_*}{R^2} \frac{4\pi}{3} \rho_{\rm s} a^3
    \end{equation}
Setting $F_{\rm D} > F_{\rm G}$, we find that the drag is only able to overcome the gravitational force for particles smaller than the maximum entrained size $a_{\mathrm{ent}}$ \citep{Facchini_16}:
\begin{equation}
    a < a_{\mathrm{ent}} = \frac{v_{\rm th}}{GM_*} \frac{\dot{M}}{4\pi \mathcal{F} \rho_{\rm s}}
    \label{eq:a_ent}
    .
\end{equation}

Any dust that is smaller than this entrained size can be blown away by the wind.
We now work out the entrained mass fraction $f_{\rm ent}$, assuming a size distribution where the number of grains of sizes between $a$ and $a+{\rm d}a$ is $n(a){\rm d}a \propto a^{-p}{\rm d}a$.
Following \citet{Haworth_18_Trappist}, we assume a MRN distribution \citep{MRN_1977} with $p=3.5$, even though such a distribution is probably only appropriate for a collisional distribution, such as fragmentation limited dust. Although \citet{Birnstiel_12} suggests that drift limited dust has a size distribution with mass more concentrated towards larger sizes i.e. $p<3.5$, in Section \ref{sec:parameters}, we confirm that in fact our results are insensitive to the value of $p$ and we may safely take $p=3.5$ as a fiducial value.

\begin{equation}
    f_{\mathrm{ent}} = \min\left[1,\left(\frac{a_{\mathrm{ent}}}{a_{\mathrm{max}}}\right)^{4-p}\right]
    .
    \label{eq:entrained_fraction}
\end{equation}

The mass in dust removed from a cell is thus $f_{\rm ent}$ times the mass in gas that is driven off in the wind multiplied by the dust-to-gas ratio $\frac{M_{\rm{d},i}}{M_i}$:
\begin{equation}
    \dot{M}_{\mathrm{},i} = f_{\mathrm{ent},i} \frac{M_{\rm{d},i}}{M_i}
    \dot{M}_{\mathrm{eff},i}
    .
\end{equation}
Although small dust could in principle be preferentially removed, the code maintains a fixed ratio of the mass contained in the larger and small grain populations.
We are able to do this because, as described above and in Section \ref{sec:parameters}, the effects of entrainment in the wind are insensitive to the grain size distribution since the majority of the dust is lost while the grains are small, at which point  all grain sizes can be removed.
A similar assumption is made by the model of \cite{Birnstiel_12} in updating the mass fraction in each population in each cell after recalculating the sizes and applying the radial drift.

\section{Gas evolution}
\begin{figure*}
    \centering
    \begin{subfigure}{0.49\linewidth}
        \centering
        \includegraphics[width=\linewidth]{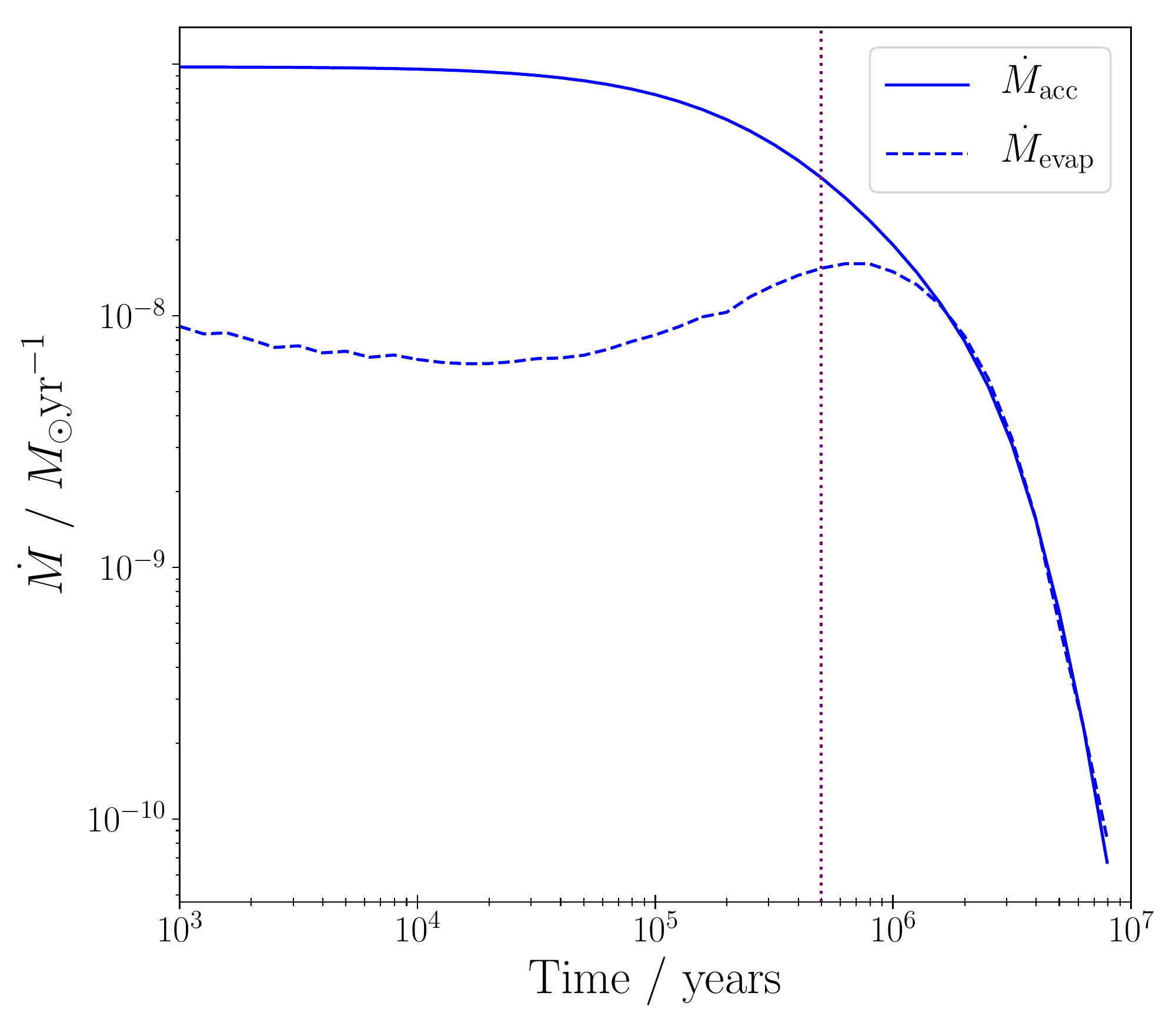}
        \caption{Initially small ($R_{\rm C} = 10~\mathrm{au}$) disc.}
        \label{fig:low_high10}
    \end{subfigure}
    \begin{subfigure}{0.49\linewidth}
        \centering
        \includegraphics[width=\linewidth]{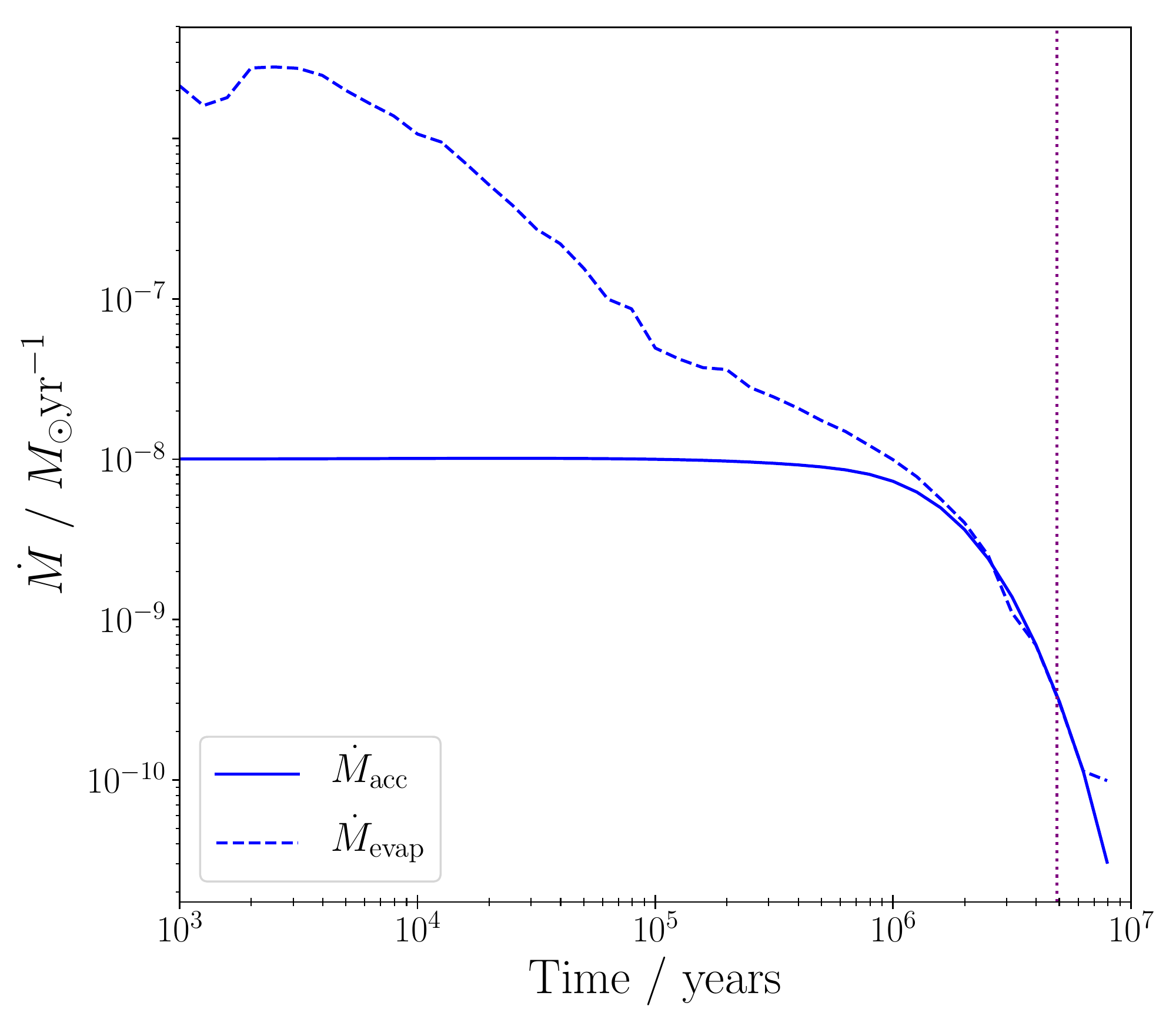}
        \caption{Initially large ($R_{\rm C} = 100~\mathrm{au}$) disc.}
        \label{fig:low_high100}
    \end{subfigure}
    \caption{The photoevaporative mass loss rate (dashed) and the accretion rate at the inner boundary (solid) as a function of age for models with $\alpha=10^{-3}$ and $\mathrm{FUV}=1000~{\rm G_0}$. 
    The vertical purple line indicates the viscous time-scale of each model at $R_{\rm C}$.}
\end{figure*}

We have run a suite of gas-only models in order to compare with previous studies, confirming that the key behaviours seen in \citet{Clarke_2007} still occur in our new model set using the updated prescriptions of \citet{Haworth_18_FRIED}. Specifically we confirm that the late time (i.e. on a few Myr) behaviour of discs with fixed FUV flux and $\alpha$ value is rather insensitive to the initial value of the scale radius, $R_{\rm C}$. 
Small discs (Fig. \ref{fig:low_high10}) start by viscously spreading until they reach the point where their outer radius shrinks with time due to photoevaporation. Conversely, an initially extended disc (Fig. \ref{fig:low_high100}) may shrink throughout its lifetime due to photoevaporation; the dominance of photoevaporation over viscous spreading is strong enough that disc shrinkage may be more or less what it would be in the absence of viscosity.
At late times, however, both simulations converge on the same tracks and the mass loss rates due to photoevaporation and viscous accretion decline in tandem as shown in Fig. \ref{fig:low_high10} and \ref{fig:low_high100} (cf. Figures 2 and 3  of \citealt{Clarke_2007}). This represents a state of self-adjustment where the evolution of the disc radius is such that the photoevaporative wind can remove the outward flux of material associated with viscous accretion on to the star (note that the disc outer edge is, in the strongly photoevaporated state, located at a small multiple of the disc half mass radius and so removal of the angular momentum from the accretion flow involves a wind mass loss rate that is similar to the accretion rate). 
 
\begin{figure}
    \centering
    \includegraphics[width=\linewidth]{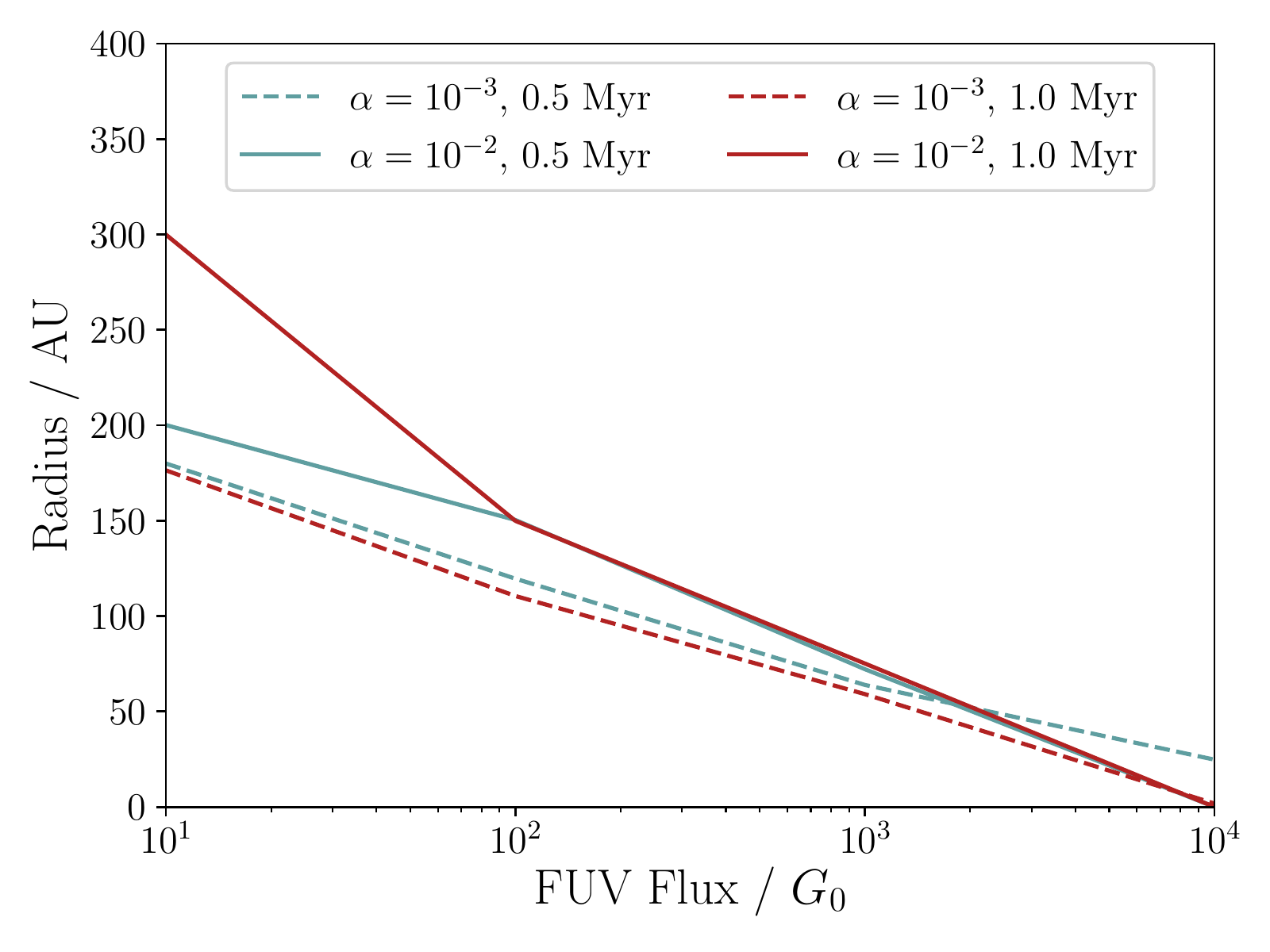}
    \caption{The disc radius as a function of the UV environment for a $\alpha = 10^{-3}$ (dashed) and $10^{-2}$ (solid) taken at $0.5$ Myr (blue) and $1$ Myr (red).}
    \label{fig:gas_radius}
\end{figure}

Secondly, as noted above,  we find that the evolution of large discs (i.e. those that shrink throughout their evolution) is rather insensitive to $\alpha$ but depends on the FUV flux, with discs shrinking more rapidly in the presence of stronger FUV fields.
Fig. \ref{fig:gas_radius} shows how as we go to higher FUV fields, the radii of the discs are increasingly convergent between models with different $\alpha$. This behaviour can be understood in terms of discs where viscous evolution is initially sub-dominant (due to the long viscous times and short photoevaporation times) and so the disc is eroded at a rate that mainly depends on the efficacy of photoevaporation, which is a very strong function of radius.
Initially small discs, conversely, start off by viscously expanding and therefore the time at which they attain maximum radius depends on $\alpha$ (and also on the FUV flux, note how for $\alpha=10^{-2}$ the radius is still increasing between $0.5$ and $1$ Myr in Fig. \ref{fig:gas_radius} at $G_0=10$ but has stalled elsewhere) since this controls how quickly they reach the point at which photoevaporation becomes competitive with viscous evolution and hence the outward expansion is halted. See \citet{Clarke_2007,Haworth_18_Trappist,Winter_2019} for further discussion of the evolution of gas in discs subject to external photoevaporation.

\section{Dust evolution in photoevaporating discs}
There are three key processes that  remove dust from the disc, thus contributing to the observed decline in  dust masses with age  \citep[e.g.][]{ALMA_Orionis}. These are: accretion following the viscous motion of the gas, radial drift due to drag from the gas, and photoevaporation due to FUV irradiation. We now assess the relative contribution of each to the dust depletion and their respective time-scales.

In this section we first examine a fiducial model of a $100~\rm{M_{J}}$, $R_{\rm C}=100~\mathrm{au}$ disc with $\alpha=10^{-3}$ around a $1~\rm{M_{\sun}}$ star in order to understand the key phases in the evolution of such discs. In the absence of photoevaporation or radial drift, dust would, in this fiducial simulation, be lost to the star by viscous accretion on a
time-scale of $10$ Myr. 
Since  photoevaporation and radial drift compete to remove dust at a different rate to the gas, we examine three scenarios in order to assess their relative importance: model RD,  with radial drift but no photoevaporation, model PE,  with photoevaporation ($1000~{\rm G_0}$) but no radial drift, and  model RDPE with both radial drift and photoevaporation.

\begin{figure*}
    \centering
    \includegraphics[width=\linewidth]{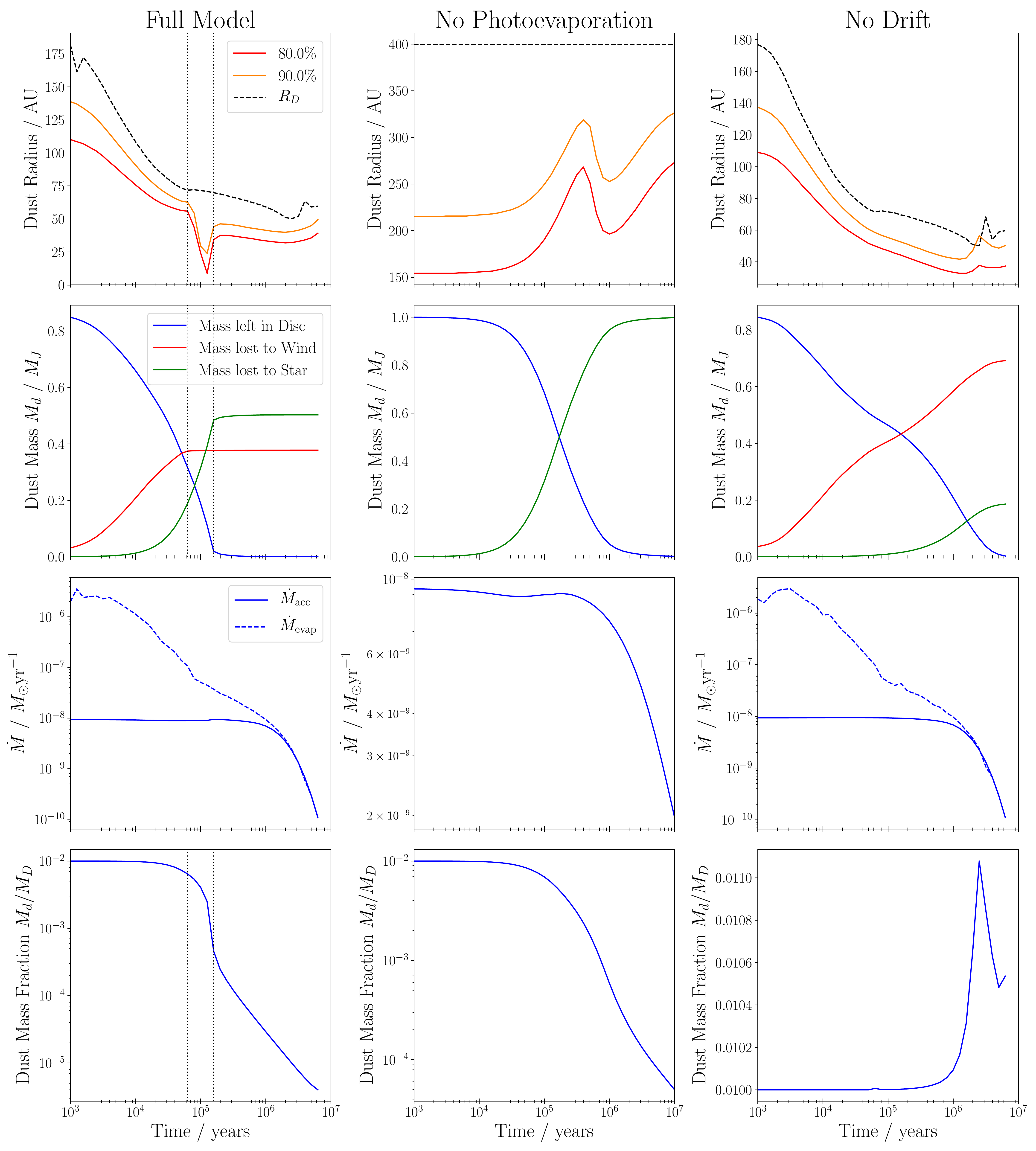}
    \caption{From left to right: Model RDPE including radial drift and photoevaporation, Model RD including radial drift but no photoevaporation and Model PE including photoevaporation but no radial drift. The panels show, from top: the radius of the 90 per cent (orange) and 80 per cent (red) percentiles of the dust mass distribution with the radius of the gas disc (black, dashed) for reference; the mass in dust left in the disc (blue), lost to the wind (red) and lost to the star (green); the global dust mass fraction; the gas mass loss rate through accretion (solid) and photoevaporation (dashed). In addition, the time of the dust at the wind base beginning to radially drift in and having cleared the disc are shown as the two black dotted lines. Note the linear scale on the bottom right plot due to the small variation in the dust mass fraction.}
    \label{fig:dust_evolution_comparison}
\end{figure*}

Fig. \ref{fig:dust_evolution_comparison} shows (from left to right) the dust evolution of Models RDPE, RD and PE, tracking the dust mass radii (defined as the radii containing 80 and 90 per cent of the dust mass at that time), the fractions of the initial dust mass that end up in different locations, the accretion and photoevaporation mass loss rates and the over-all dust to gas ratio of the remaining disc, $M_{\rm d}/M_{\rm D}$.

\begin{figure*}
    \centering
    \includegraphics[width=\linewidth]{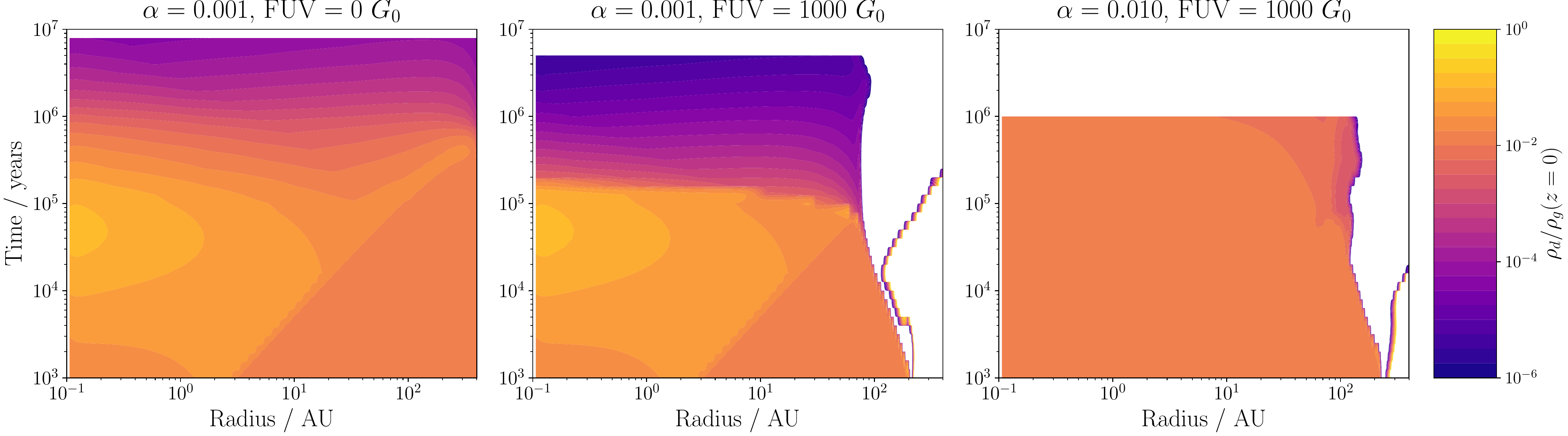}
    \caption{Contour plots of the midplane dust-to-gas ratio $\rho_{\rm d}/\rho_{\rm g}(z=0)$ for L: a model without photoevaporation, C: a model with an FUV flux of $1000~{\rm G_0}$, R: the same but with high $\alpha$. Light orange to yellow colours represent enhancement over $0.01$ while purple and blue colours represent a lower ratio.}
    \label{fig:d2g_vsCarrera}
\end{figure*}

In the absence of photoevaporation, the dust, which starts off small and hence well entrained everywhere, grows quickly in the inner disc  and starts radially drifting after around $10^4~\mathrm{yr}$. Since this results in the innermost parts of the dust distribution being lost to the star, the 80th and 90th percentiles initially move outward. Radial drift causes the mass of dust to decline rapidly around $0.1-0.2~\mathrm{Myr}$, though the small dust component provides a reservoir that lasts for $3.2~\mathrm{Myr}$. Once most of the mass in the disc is drifting, the percentiles move inwards \citep{Rosotti_2019b}.
The significant fall in  the global dust mass fraction is a well known problem for models incorporating radial drift \citep{Takeuchi_2005,Brauer_2008,Birnstiel_12,Pinilla_2012} unless the disc's turbulent  viscosity is high enough for fragmentation to keep the bulk of the dust grains small and  hence well coupled to the gas. The left hand panel of Fig. \ref{fig:d2g_vsCarrera} illustrates the evolution of the radial profile of mid-plane dust to gas ratio\footnote{Note that the calculation of the mid-plane dust to gas ratio from the vertically averaged quantity accounts for vertical settling of dust: see Youdin \& Lithwick 2007, Birnstiel et al 2012. } in the absence of photoevaporation. 

Conversely, in model PE, the dust radius follows the gas radius which, in this simulation, falls monotonically after the first $\sim 10^3$ years due to photoevaporation. The preferential removal of gas compared with dust is indicated by a small rise in the dust to gas ratio in the residual disc. This simulation (in which dust is not allowed to migrate) represents the case of maximal entrainment of dust in the wind, with the total dust lost to the wind exceeding that accreted on to the star by a factor $\sim 3-4$.  The time-scale for dust depletion is however rather longer than in model RD.

\begin{figure*}
    \centering
    \includegraphics[width=\linewidth]{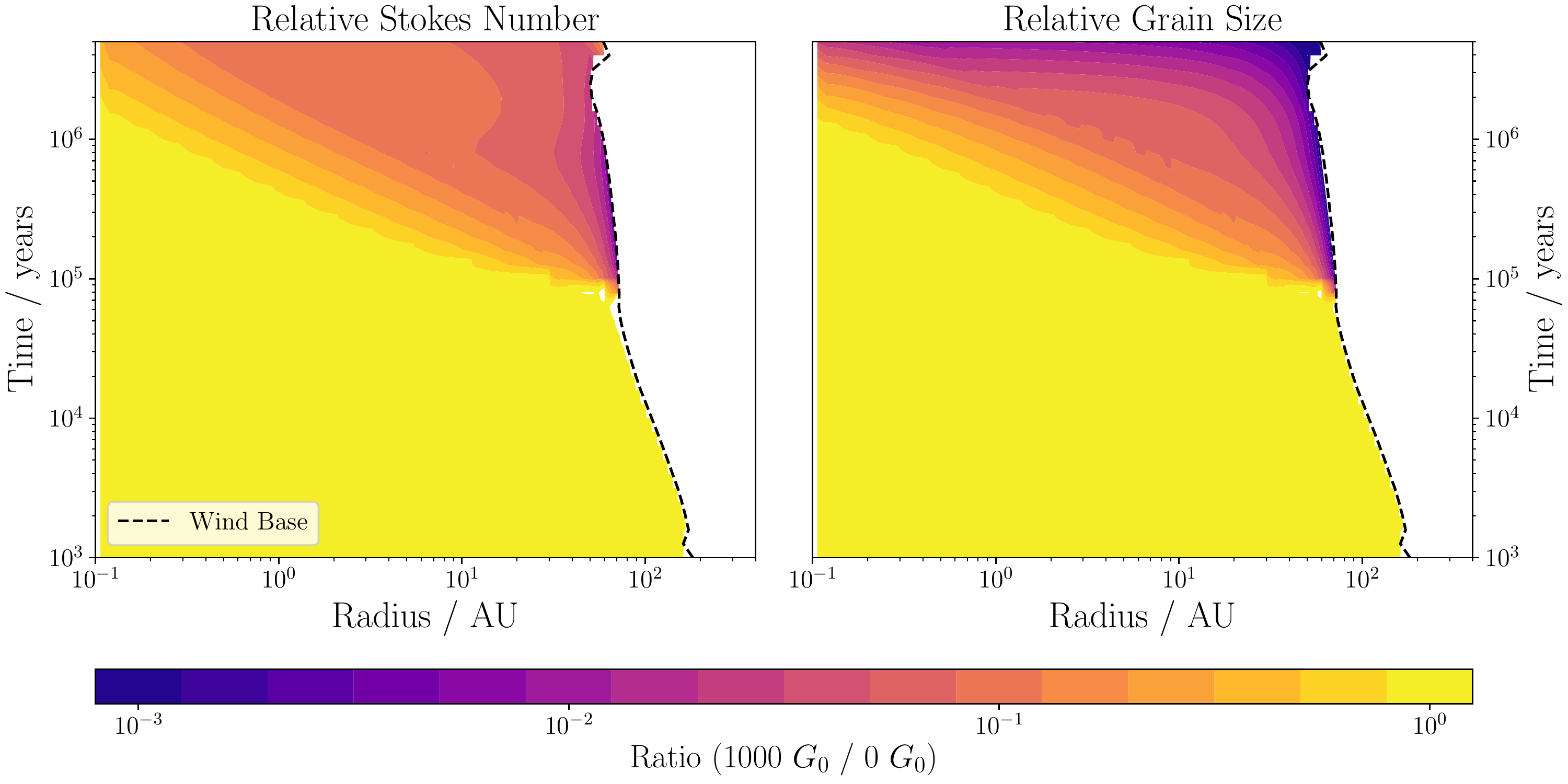}
    \caption{Effect of Photoevaporation on L: the maximum Stokes number, R: the maximum dust grain size. Shown in each panel are the values for the model with $G_0 = 1000$ normalised by the values for the model with no photoevaporation. The dashed black line is the radius at the base of the photoevaporative wind. After $0.1-1~\mathrm{Myr}$, consistently low Stokes numbers are seen. These correlate to the onset of lower sizes of dust. At later times, the sizes are further suppressed due to lowered gas surface densities.}
    \label{fig:size_suppression}
\end{figure*}

When both radial drift and photoevaporation are included in model RDPE, we see quite different behaviour. The loss of dust to the wind ceases sharply around $0.06~\mathrm{Myr}$ - the red line in the second left hand panel of  Fig. \ref{fig:dust_evolution_comparison} flattens out. This time is marked with the first black vertical dotted line at around $0.06~\mathrm{Myr}$. After this point, the dust in the disc is rapidly depleted by radial drift on to the star within a further  $0.1-0.2~\mathrm{Myr}$, marked with the second dotted line. The central panel of Fig. \ref{fig:d2g_vsCarrera} depicts the evolution of the profile of mid-plane dust to gas ratio in this case: comparison with the left hand panel (radial drift, no photoevaporation) demonstrates that the dust to gas ratio falls significantly more rapidly when photoevaporation is included because the previous erosion of the outer dust disc by photoevaporation means that there is a reduced reservoir for re-supply of dust by radial inflow at the point in the evolution ($\sim 0.1$ Myr) when radial drift becomes important. This lower dust to gas ratio in the presence of photoevaporation reduces particle growth and hence both the maximum grain size and the associated Stokes number are lower than in the case with radial drift alone (see Fig. \ref{fig:size_suppression}).

The onset of radial draining of the large dust component   can be understood by considering the relative speeds of radial drift and viscous motions.
Following \citet{Birnstiel_12}, we may define $St_{\rm eq}$ as the Stokes number where the radial drift velocity of the dust equals the viscous speed of the material (whether inwardly or outwardly directed). The gradient of the surface density enters the latter through the gradient of the torque and the former through the gradient of the pressure. Given our imposed models for the temperature,
\begin{equation}
    St_{\rm eq} = 3 \alpha \left|\frac{\frac{3}{2} + \frac{{\rm d}\ln\Sigma}{{\rm d}\ln R}}{\frac{7}{4} - \frac{{\rm d}\ln\Sigma}{{\rm d}\ln R}}\right|
    \label{eq:St_eq}
    .
\end{equation}
where $\Sigma \propto R^{-1}$ in the inner disc, $St_{\rm eq} \to \frac{6}{11} \alpha $ \citep{Birnstiel_12}. Here both velocities are inwards, but we can still use $St_{\rm eq}$ to quantify whether the radial drift or viscous dynamics are dominant.
In the outer disc, the viscous velocity is outwards and $ |{\rm d}\ln\Sigma/{\rm d}\ln R| \gg 1$, thus $ St_{\rm eq} \to 3 \alpha$.
Note that somewhere in the middle, where ${\rm d}\ln\Sigma/{\rm d}\ln R \to -3/2$, we transition from viscously accreting to spreading regimes and $St_{\rm eq} \to 0$ (because, where the viscous speed is close to zero, radial drift is 
dominant even for tightly coupled grains).

If dust at the disc edge grows beyond $St_{\rm eq} \sim 3 \alpha$, its net velocity is inwards - no amount of viscous spreading can overcome the radial drift. The viscous spreading at the edge of the disc can no longer replenish the dust to the wind base - the dust escapes the wind through radial drift to smaller radii.
This effectively shuts off the loss of dust to the wind.

At this point, since the radial drift time-scale at the maximum grain size is less than the viscous time-scale, this  dust rapidly depletes on to the star, leaving only the residual `small dust' component which remains well coupled to the gas (and whose evolution dominates the evolution of the dust radii, top row of Fig. \ref{fig:dust_evolution_comparison}, at late times.

Note that unlike the case with no radial drift, the global dust mass fraction decreases steadily with time. By the time that the dust is not perfectly entrained in the flow, significant amounts are already being lost to radial drift.
This suggests that photoevaporation is unlikely to resolve the discrepancy between our radial drift model, and the increased dust mass fractions observed by e.g. \citet{ALMA_Lupus}.

Finally we stress that the combination of photoevaporation and radial drift can shorten the lifetime of  dust in the disc by over an order of magnitude compared with models that involve only photoevaporation or only radial drift.
We discuss the implications of this result for the demographics of observed discs (Section \ref{sec:obs}) and for planet formation (Section \ref{sec:planet_form}).

\section{Parameter exploration}
\label{sec:parameters}

We now examine a grid of model results, focusing on the eventual destination of the dust (i.e. accretion onto the star or mass loss in the wind) and associated time-scales for disc depletion. 

\begin{figure*}
    \centering
    \includegraphics[width=\linewidth]{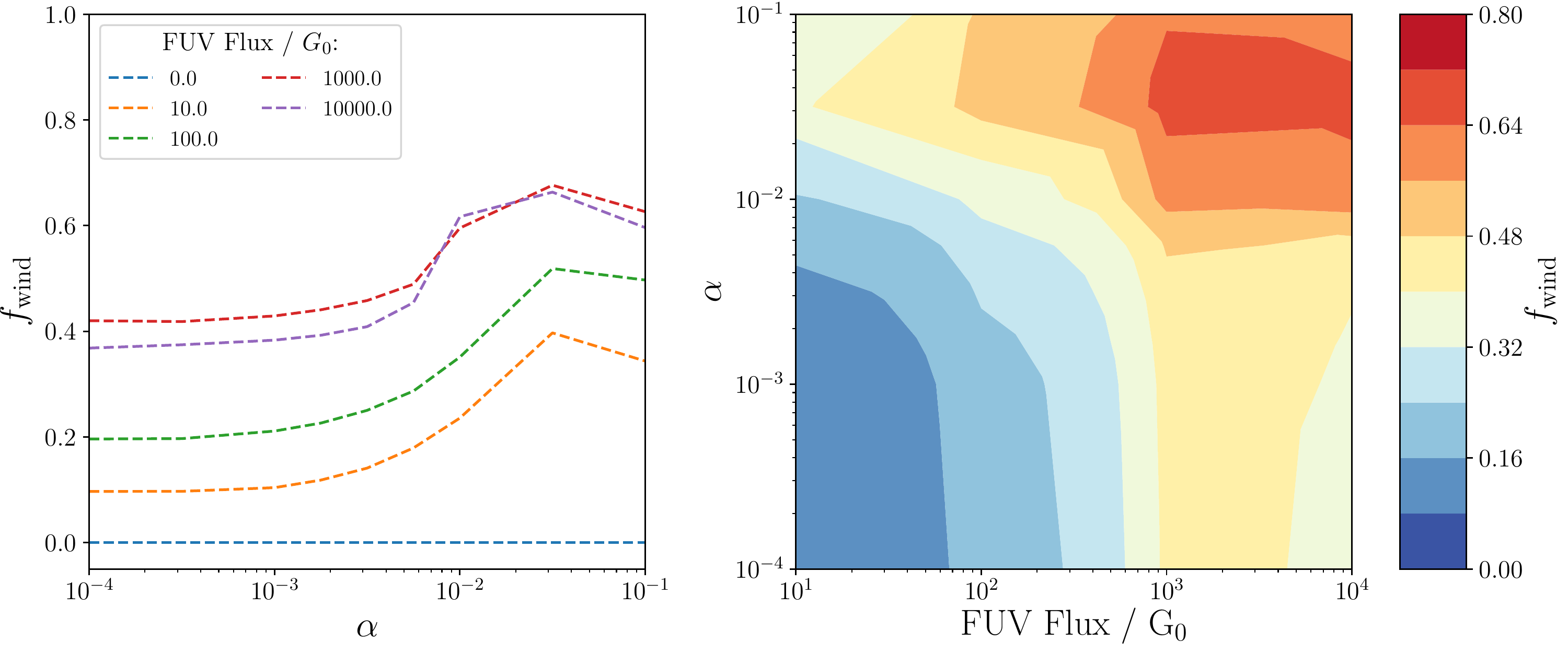}
    \caption{Effect of viscosity and FUV flux on dust loss to wind. L: $f_{\rm wind}$ as a function of $\alpha$. Each colour represents a different FUV flux. R: the plane of $\alpha$ and FUV flux, showing strong dependence on FUV flux at all $\alpha$. The colour represents  $f_{\rm wind}$ values running from $0.00$ (blue) to $0.80$ (red).}
    \label{fig:f_wind_dual}
\end{figure*}

The overwhelmingly important parameter in determining the fraction of the dust mass lost to the wind and the associated time-scales for dust depletion is the ultra-violet flux level (see right hand panel of Fig. \ref{fig:f_wind_dual} and Table \ref{tab:UV_flux}).
This result is readily explicable in terms of the higher temperatures and greater penetration depth in the case of more intense FUV fields. Nevertheless it is notable that the maximum fraction of dust lost to the wind is $\sim 65$ per cent and thus the wind is never the overwhelmingly dominant destination for the dust. The division of dust between accretion on to the star and loss in the wind approaches $50:50$ for $G_0 \sim 1000$ but does not rise steeply for higher fluxes.
\footnote{In fact, for most viscosities, the fraction of dust lost to the wind decreases slightly at the highest modelled FUV fluxes of $10^4~G_0$. This is because at larger radii and high $G_0$, the mass loss rates in the FRIED grid are not monotonic with the FUV flux (see Figure 3 of \citet{Haworth_18_FRIED}), which  these authors attribute to the critical radius of the wind coinciding with the hydrogen ionisation front. This slight decline does not affect the monotonic decrease of the depletion timescale with $G_0$ in Table \ref{tab:UV_flux} but, since much of the dust mass is lost at early times when the disc is large, can slightly decrease the fraction of dust lost in the wind.}

Another readily explicable result is that the importance of photoevaporation (both in terms of fraction of dust leaving in the wind and a short depletion time-scale) is somewhat greater for lower mass stars (Table \ref{tab:m_star}). This simply derives from the shallower potential and hence lower requirement on the escape velocity for low mass stars.
 
The fraction of the dust that is lost in the wind increases with the initial disc scale radius (Table \ref{tab:scale_radii}). This is as expected since for smaller discs the time-scale for growth and radial migration of dust grains is smaller while photoevaporation rates are lower from locations deeper within the stellar potential. The depletion time is less sensitive to initial scale radius because the interplay between photoevaporation and viscous evolution leads to a  convergence in the evolution of disc radius at late times.
  
By contrast the initial disc mass has almost no effect on the fate of the dust  and the influence on the time-scale (which is defined as time required to attain a fixed fraction of the initial dust mass) is likewise weak (Table \ref{tab:dust_mass}). In more massive discs the dust has to grow to larger size scales before it undergoes strong radial drift and this very mildly favours dust loss in the wind. 
   
   \begin{figure*}
    \centering
    \includegraphics[width=\linewidth]{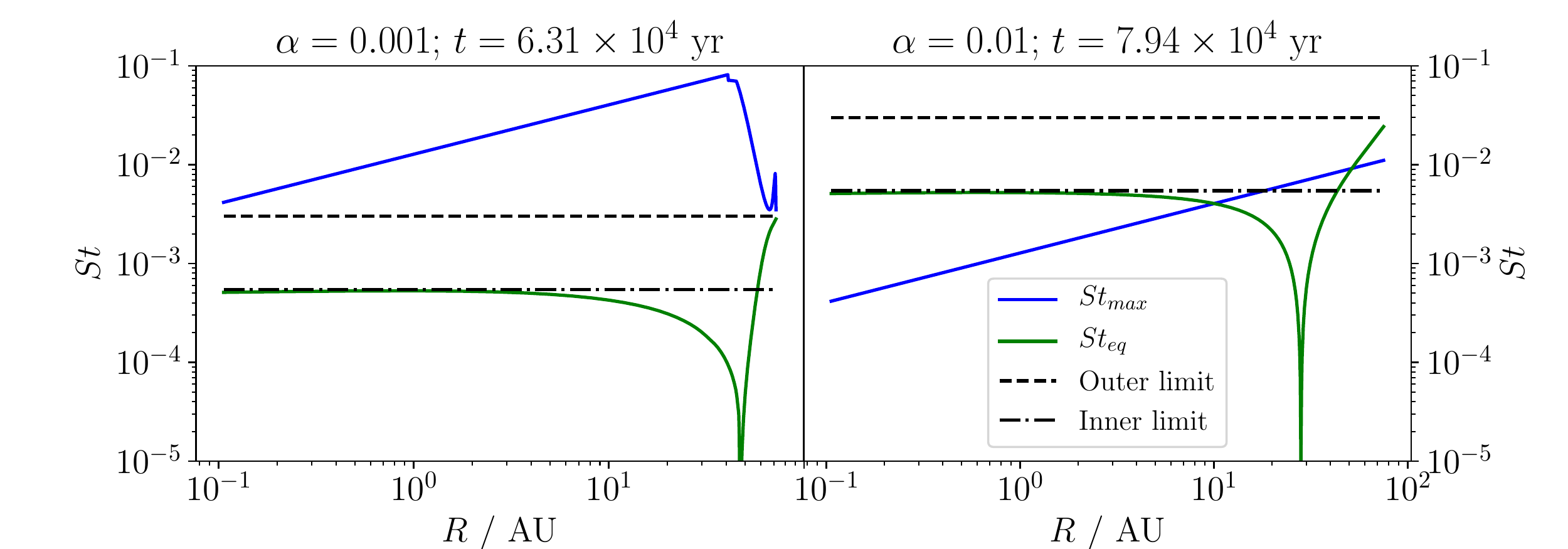}
    \caption{The Stokes number of the maximum dust size, $St_{\mathrm{max}}$ (blue) compared to that for which the background viscous advection velocity and radial drift velocity are equal, $St_{\rm eq}$ (green) given by Equation \ref{eq:St_eq}. The black dashed/dashed-dotted lines represent the limits of this expression. The left hand panel shows the low viscosity $\alpha=10^{-3}$ model after $6.31\times10^4~\mathrm{yr}$ and the right shows the high viscosity $\alpha=10^{-2}$ model after $7.94\times10^4~\mathrm{yr}$ - these were the snapshots with the highest value of $St_{\mathrm{max}}$ at the disc edge.}
    \label{fig:Stokes_optimum}
\end{figure*}

There is generally a mild positive relationship between dust loss in the wind and the value of $\alpha$ (see Table  \ref{tab:viscosity} and Fig. \ref{fig:f_wind_dual}). This partly reflects the fact that at higher viscosity the disc gas is more vigorously fed into the wind-launching zone but also because the dust in this region is less efficiently flushed inwards by radial drift.
The latter contribution is illustrated in Fig. \ref{fig:Stokes_optimum} which plots the radial profile of the Stokes number at the point that the Stokes number at the wind base is a maximum for the cases of simulations with $\alpha = 0.001$ and $0.01$. Fig. \ref{fig:Stokes_optimum} also depicts, in each case, the profile of the Stokes number for which the inward radial drift speed is equal to the magnitude of the  viscous speed in the disc ($St_{\rm eq}$, as defined by Eq \ref{eq:St_eq}). Clearly in the low viscosity case the dust can drift inwards on less than the viscous time-scale, whereas for high viscosity, fragmentation keeps the grains small and the radial motion of the dust is well coupled to the viscous flow of the gas. The right hand panel of Fig. \ref{fig:d2g_vsCarrera} demonstrates the milder evolution of the dust to gas ratio in the case of a higher viscosity model.
    
For low-medium $\alpha$ values, the dependence on $\alpha$ is thus very weak because the dust takes the same time to grow to the point  that radial drift removes it from the wind base. At higher $\alpha$, the reduction in grain sizes due to fragmentation starts to suppress radial drift and hence the dependence on $\alpha$ becomes steeper. Only at the highest values, above those typically inferred from observations \citep[e.g.][]{Rafikov_2017}, does the fraction of dust lost start to decrease again. This is  because the grains are already tightly coupled to the gas, thus limiting the improvement possible, but viscous accretion becomes a competitive sink for the material. In line with this, we see that the depletion timescale first rises, as radial drift becomes less effective, and decreases again once viscous accretion becomes effective.
    
We also vary the parameter $\gamma$ that determines the initial density profile, and $q$ that determines the temperature profile, trying a steeper profile for each, and starting both in and out of the steady state (Table \ref{tab:profiles}).
Changing the temperature profile changes the viscosity, so so long as drift dominates the dust motions, changing the temperature profile alone has little effect on the dust fates and lifetimes.
The slope of the surface density has a larger effect - for $\gamma=3/2$, the material is more centrally concentrated and so is depleted by drift more rapidly, thus reducing the efficacy of the wind at removing the dust. When $\gamma=3/4$ is adopted in line with the steady state solution for $q=3/4$, the material is now less centrally concentrated, aiding removal by the wind. The depletion time-scale is less-affected, in all cases lying within $0.2 \pm 0.05~\rm Myr$.
    
Finally, we check some properties of the dust model.
At both high and low $\alpha$, we calculated $f_{\rm ent}$ (Equation \ref{eq:entrained_fraction}) using a more top heavy dust size distribution with slope $p=2.5$. We find that the entrained fractions change by a fraction of a per cent compared to those in Table \ref{tab:viscosity} - so the results are robust against uncertainties in the grain size distribution.
The reason for this insensitivity is that at low $\alpha$, dust loss is curtailed not by grains being too large to be entrained in the wind but through being subject to efficient radial drift. Thus dust loss occurs only while grains are still small and $f_{\rm ent}=1$.
At high $\alpha$, the dust loss is ultimately controlled by the supply of dust to the wind base by viscous spreading, again making the amount of mass entrained insensitive to the grain size distribution.
We also test the impact of starting all the dust in our fiducial model at its limiting size, rather than letting it grow from the monomer size of $0.1~\rm\mu m$. This does not change the qualitative behaviour, but since previously dust was largely lost during the growth phase before radial drift set in,  $f_{\rm wind}$ reduces by a factor $\sim2$ from $0.43$ to $0.18$.
If dust grains are processed in the formation of the disc and hence start out larger than we assume, we should thus expect $f_{\rm wind}$ to be reduced by a factor $\lesssim 2$.
    
 \begin{table}
    Key model outcomes. $f_{\rm wind}$ and $f_{\rm star}$ represent the mass fraction of the dust lost to the wind and the fraction of the dust lost onto the star through viscous accretion or radial drift respectively. $t_{\rm dep}$ is the time after which only $1\%$ of the initial dust mass remains.
     \caption{Models with varying FUV flux.}
        \label{tab:UV_flux}
        \begin{tabular}{c|c|c|c} \hline
            FUV Flux ($\mathrm{G_0}$) & $f_{\rm wind}$ & $f_{\rm star}$ & $t_{\rm dep}$ ($\mathrm{Myr}$) \\ \hline
            0    &   0.000   & 0.998  &   3.181   \\
            10   &   0.104   & 0.896  &   1.026   \\
            100  &   0.211   & 0.789  &   0.604   \\
            1000 &   0.429   & 0.571  &   0.219   \\
            10000&   0.383   & 0.617  &   0.146   \\ \hline
        \end{tabular}
        \\
        \caption{Models with varying stellar mass.}
        \label{tab:m_star}
        \begin{tabular}{c|c|c|c} \hline
            $M_*$ ($M_{\sun}$) & $f_{\rm wind}$ & $f_{\rm star}$ & $t_{\rm dep}$ ($\mathrm{Myr}$) \\ \hline
            0.5  &   0.557   & 0.443  &   0.122    \\
            1.0  &   0.429   & 0.571  &   0.219    \\
            1.9  &   0.259   & 0.741  &   0.376    \\ \hline
        \end{tabular}
    \\
        \caption{Models with varying scale radius.}
        \label{tab:scale_radii}
        \begin{tabular}{c|c|c|c} \hline
            $R_{\rm C}$ ($\mathrm{au}$) & $f_{\rm wind}$ & $f_{\rm star}$ & $t_{\rm dep}$ ($\mathrm{Myr}$) \\ \hline
            10  &   0.003   & 0.997  &   0.151    \\
            30  &   0.096   & 0.904  &   0.250    \\
            100 &   0.429   & 0.571  &   0.219    \\
            300 &   0.584   & 0.416  &   0.190    \\ \hline
        \end{tabular}
        \\
        \caption{Models with varying initial disc mass.}
        \label{tab:dust_mass}
        \begin{tabular}{c|c|c|c} \hline
            $M_{\rm D}$ ($\mathrm{M_{\rm J}}$) & $f_{\rm wind}$ & $f_{\rm star}$ & $t_{\rm dep}$ ($\mathrm{Myr}$) \\ \hline
            100  &   0.429   & 0.571  &   0.219    \\
            30   &   0.394   & 0.606  &   0.193    \\
            10   &   0.362   & 0.638  &   0.183    \\
            3    &   0.379   & 0.621  &   0.208    \\
            1    &   0.337   & 0.663  &   0.211    \\ \hline
        \end{tabular}
    \\
     \caption{Models with varying viscosity ($\alpha$).}
        \label{tab:viscosity}
        \begin{tabular}{c|c|c|c} \hline
            $\alpha$ & $f_{\rm wind}$ & $f_{\rm star}$ & $t_{\rm dep}$ ($\mathrm{Myr}$) \\ \hline
            $1\times10^{-4}$  &   0.420   & 0.580  &   0.213    \\
            $3\times10^{-4}$  &   0.418   & 0.582  &   0.214    \\
            $1\times10^{-3}$  &   0.429   & 0.571  &   0.219    \\
            $3\times10^{-3}$  &   0.458   & 0.542  &   0.325    \\
            $1\times10^{-2}$  &   0.595   & 0.404  &   0.659    \\
            $3\times10^{-2}$  &   0.676   & 0.323  &   0.271    \\
            $1\times10^{-1}$  &   0.626   & 0.374  &   0.113    \\ \hline
        \end{tabular}
    \\
     \caption{Models with varying power law indices for temperature ($q$) and initial surface density ($\gamma$): see equations \ref{eq:T_law} and \ref{eq:LBP_profile}}
        \label{tab:profiles}
        \begin{tabular}{c|c|c|c|c|c} \hline
            $q$ & $\gamma$ & Initially Steady State? & $f_{\rm wind}$ & $f_{\rm star}$ & $t_{\rm dep}$ ($\mathrm{Myr}$) \\ \hline
            $0.5$  &    $1$     &   Y   &   0.429   & 0.571  &   0.219    \\
            $0.5$  &    $3/2$   &   N   &   0.276   & 0.724  &   0.198    \\
            $0$    &    $3/2$   &   Y   &   0.302   & 0.697  &   0.157    \\
            $0.75$ &    $1$     &   N   &   0.424   & 0.574  &   0.230    \\
            $0.75$ &    $3/4$   &   Y   &   0.480   & 0.520  &   0.231    \\ \hline
        \end{tabular}
        \end {table}
        
\section{Discussion: predictions for disc demographics}
\label{sec:obs}
The suite of results presented above, as summarised in Table 1,  implies that photoevaporation clearly reduces the lifetime of both the gas and the dust in protoplanetary discs. The time-scale on which $99$ per cent of the dust is lost from the disc is reduced as a result of photoevaporation even when (as in the majority of cases) only a relatively  minor component  of the dust (few 10s of per cent) is actually removed in the wind (see Fig. \ref{fig:f_wind_dual}).
The main effect that causes photoevaporation to reduce the dust lifetime is that the wind's early removal of dust from the outer disc then prevents later replenishment of the inner disc as it becomes dust depleted on account of  radial drift.
    
Observational data on the demographics of protoplanetary discs provides information on not only the average lifetimes of discs \citep{Haisch_2001,INgleby_2012,Ribas_2015} but also on the relative lifetimes of the dust and gas components (e.g. \citet{Fedele10}). In general there is a reasonably strong correlation between those stars that exhibit a near infrared excess from warm dust and those that manifest accretion on to the star.  \citet{Fedele10} found similar time-scales for the decline of dust and gas, arguing for a slightly longer duration of the infrared excess phase      compared with the lifetime over which accretion proceeds at a detectable level (i.e. in excess of $10^{-11}~\rm{M_{\sun}}$ yr$^{-1}$).
     
\begin{figure}
    \centering
    \includegraphics[width=\linewidth]{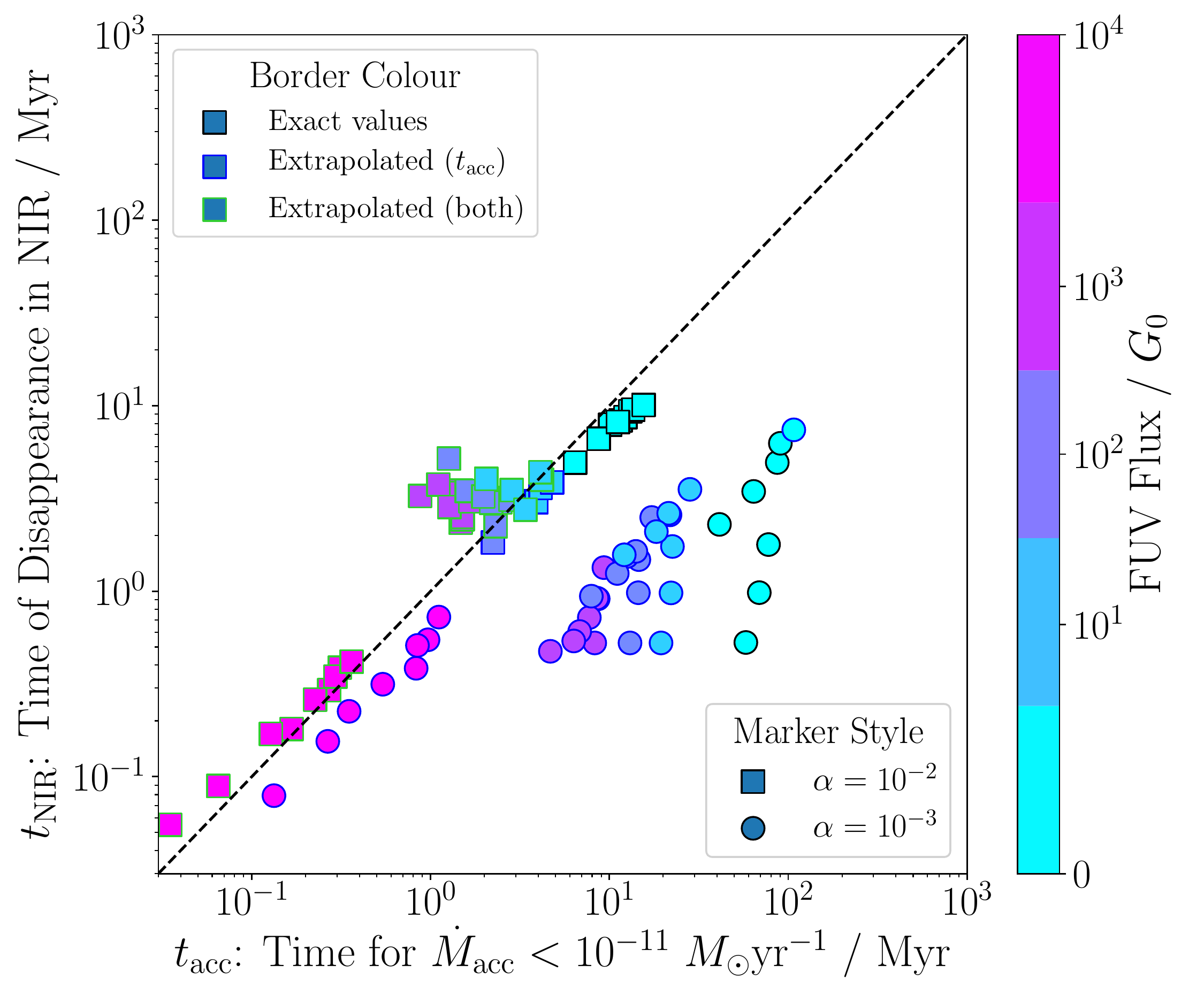}
    \caption{Suite of model results in the plane of accretion lifetime versus lifetime of near infrared excess. The symbol colour denotes FUV flux while the symbol shape differentiates simulations with $\alpha = 10^{-2}$ (square) and $10^{-3}$ (circle). The black borders are directly obtained from the simulations whereas the blue and green borders denote models where an extrapolation is involved in the accretion rate and both the accretion rate and $2~\mu$m optical depth respectively (see text). The dashed line indicates equality of time-scales.}
    \label{dustgasl}
\end{figure}

We have run a large suite of models with a range of initial disc masses and scale radii subjected to varying levels of ultraviolet background radiation at both high and low viscosities, all around a $1~\rm{M_{\sun}}$ star. Fig. \ref{dustgasl} plots the results of these models in the plane of accretion lifetime against near infrared excess lifetime where the former denotes the time over which accretion on to the star exceeds $10^{-11}~\rm{M_{\sun}}$ yr$^{-1}$ and the latter the lifetime over which the disc is optically thick at $2~ \mu$m at a radius of $1$ au
\footnote{The optical depth is computed as in \citet{Rosotti_2019a,Rosotti_2019b}, i.e. using the opacity of \citet{Tazzari_2016}}.
Note that we only pursue the calculations to the point where the photoevaporation rate declines to the minimum value covered by the FRIED grid and that this can occur before the above criteria are satisfied. In such cases we linearly extrapolate the decline in $2~\mu$m optical depth and accretion rate (considering log accretion rate and log $2~\mu$m optical depth versus linear time and log time respectively) in order to estimate the dust and gas lifetimes. The extrapolated lifetimes generally exceed the point at which the photoevaporation declines to the minimum grid value by a factor of order unity.  

It can be seen from Fig. \ref{dustgasl} that, as expected, lifetimes in both dust and gas decline as the FUV background is increased. The very short lifetimes at high flux levels (e.g. $10^4~{\rm G_0}$ as in the core of the Orion Nebula Cluster) are well known and are the origin of the `proplyd lifetime problem' \citep{Henney_1999}
\footnote{Note that while we present the analysis here in terms of the disappearance of the near infrared excess emission, which is a widely measured quantity, we have also examined the decline in submm flux. Our results confirm that even initially massive discs, when exposed to the FUV levels experienced in the core of the Orion Nebula Cluster, are expected to remain above the detection threshold of recent high sensitivity ALMA surveys of the region \citep{Eisner18} for less than a Myr. Given the large fraction of discs that are detected at this FUV level, this strengthens the argument that the disc sources in the core of the Orion Nebula Cluster must have been exposed to the current FUV field for a relatively short period in the past and that the proplyd phenomenon is likely to be short lived.}.
At flux levels above $\sim 1000~{\rm G_0}$ the dust and gas lifetimes are comparable. This means that dynamical studies in highly irradiated environments  (e.g. \citet{Scally}  for the ONC or \citet{Winter_2019} for Cygnus OB2) which measure disc lifetimes from gas-only calculations can still be compared with dust based diagnostics. On the other hand when the FUV flux is below $\sim 1000~{\rm G_0}$ the relationship between dust and gas lifetime depends on the value of the viscous $\alpha$ parameter. Broadly speaking this is because at high $\alpha$ $(10^{-2})$, turbulent fragmentation keeps grains small and well coupled to the gas and thus the lifetimes in the two diagnostics are comparable. When $\alpha$ is reduced by an order of magnitude, the gas lifetime increases since, regardless of the strength of the FUV field, viscous evolution is the main agent of disc clearing. On the other hand, the lowered turbulence allows dust to grow to sizes where it is subject to strong radial drift and hence the dust lifetime decreases. Thus lowering $\alpha$ results in models moving towards the lower right of the plot where the gas lifetime substantially exceeds the lifetime of the dust. 
  
It is interesting that, while the absolute lifetime is a strong function of FUV flux, the relationship between dust and gas lifetimes is predominantly controlled by $\alpha$. At face value the observed near equality of dust and gas lifetimes argues for the higher $\alpha$ value; we emphasise that the magnitude of this problem (the high ratio of gas to dust lifetime at low $\alpha$) is unaffected by the level of FUV flux except in regions with the  highest background levels.
We note however, that the NIR lifetime and accreting lifetime trace the presence of the dust and the dynamics of the gas in the \textit{inner disc}, so other effects - such as gap opening by planets, which can disrupt the supply of material to the inner disc \citep[e.g.][]{Armitage_1999}, or internal photoevaporation, which may lead to photoevaporation-starved accretion and gap opening \citep{Drake_2009,Owen_2011} - may instead impart a similar lifetime on both observables.
   
\begin{figure*}
    \centering
    \includegraphics[width=\linewidth]{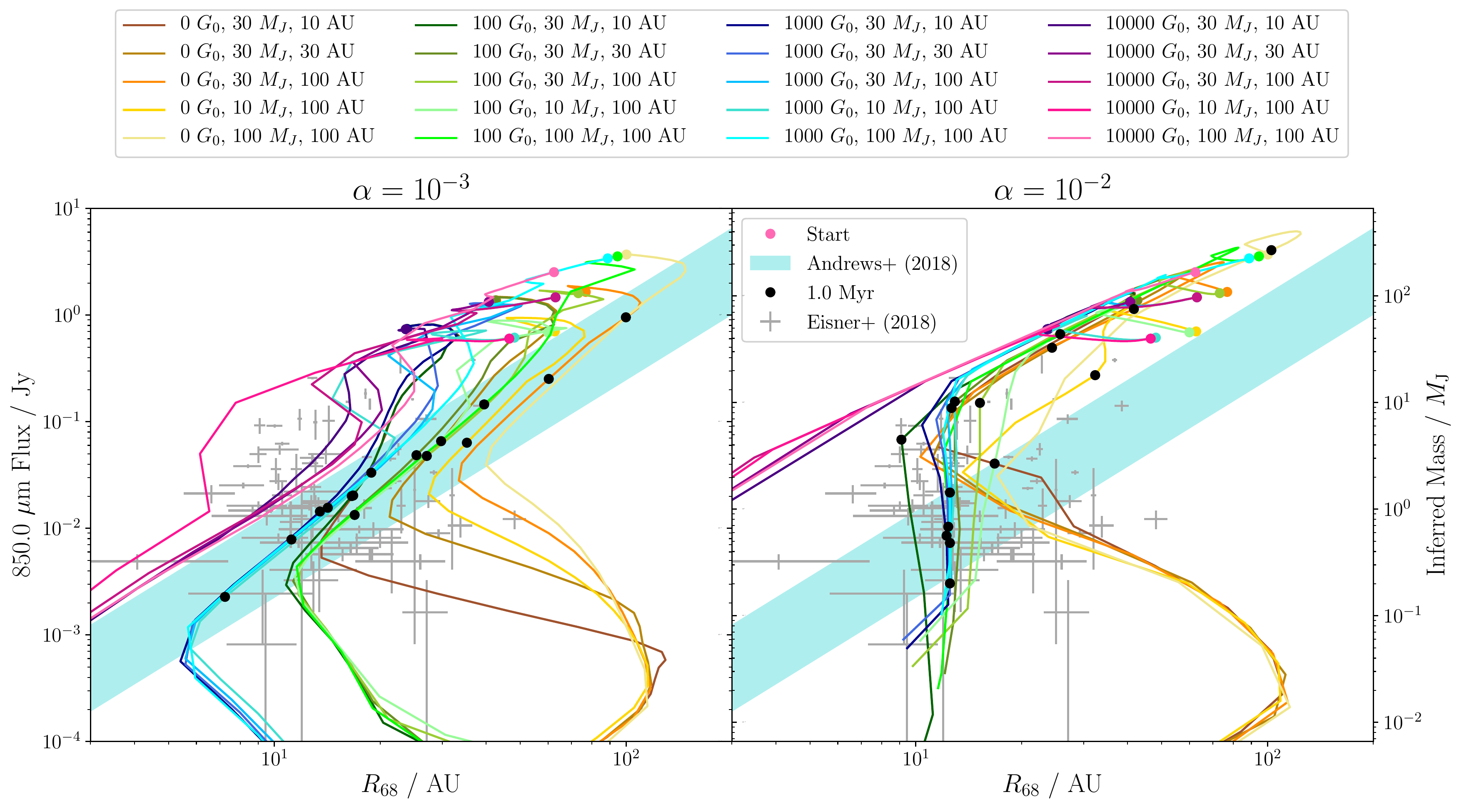}
    \caption{Trajectories of models in the plane of disc flux $68$th centile radius versus flux (at a distance of $140$ pc), both evaluated at $850~\mu$m. Left hand panel: $\alpha = 10^{-3}$, Right hand panel: $\alpha = 10^{-2}$. A black dot on each model locus indicates the properties at an age of $1$ Myr. Observational data is from \citet{Eisner18} in the case of discs residing in the highly irradiated environment of the Orion Nebula Cluster (with fluxes rescaled to a distance of $140$ pc). The blue band is the mean relation and $\pm 1 \sigma$ extent of data assembled in \citet{Tripathi}, \citet{Andrews_2018} for regions with low photoevaporation levels. On the right hand scale we indicate the total disc mass that an observer would infer if they made canonical assumptions about the dust to gas ratio and opacity (see text): we enphasise that this scale is merely given to ease comparison with observed systems and does {\it not} correspond to the actual disc masses in the model, on account of the spatially variable opacity and dust to gas ratio in our modeling.}
    \label{fig:flux-rad}
\end{figure*}
    
We have also examined the effect of photoevaporation on disc fluxes and radii at submm wavelengths. Here we follow the observational study of \citet{Tripathi} by  plotting model trajectories in the plane of $R_{68}$ (the radius enclosing $68$ per cent of the flux at $850~\mu$m) versus the corresponding flux (assuming that all sources are at a distance of $140$ pc).
On the right hand scale we  indicate the  inferred disc mass based on calculating the dust masses from the mm fluxes following the commonly used prescription of \citet{Hildebrand_1983,Beckwith_1990} and further assuming a canonical gas to dust ratio of 100 to get a total mass; since in our models we observe severe dust depletion due to radial drift, and use different, spatially variable opacities and temperatures, these do  not necessarily reflect the actual  disc masses.
In the absence of photoevaporation, the model trajectories reproduce the results of \citet{Rosotti_2019a}. Sources evolve towards lower fluxes and radii with time, an effect that can be understood in terms of the inward migration of the radius (termed $R_{\rm cliff}$ by \citet{Rosotti_2019a}) outside which the grains are sufficiently small ($< 0.1$ mm) so as to present a strongly reduced opacity at submm wavelengths.
As discussed by \citet{Rosotti_2019a}, the trajectories differ in detail depending on whether the maximum grain size is set predominantly by radial drift (low $\alpha$) or fragmentation (high $\alpha$). \citet{Rosotti_2019a} argue that the observational data is better reproduced by the former tracks and indeed the trajectory of the `drift locus' is well aligned with the mean observed relationship (\citealt{Tripathi}, \citealt{Andrews_2018}: we indicate the $\pm 1 \sigma$ interval of the observed data from these surveys as the blue band in Fig. \ref{fig:flux-rad}).
The only new element in our calculations (in the absence of photoevaporation) compared with those presented in \citet{Rosotti_2019a} is that we have pursued the evolution for longer and see that at late times the tracks move towards larger $R_{68}$ again. This is because $R_{\rm cliff}$ becomes sufficiently small that it moves within $R_{68}$ as  the flux originating from the low opacity region outside  $R_{\rm cliff}$ becomes significant. In the case of low $\alpha$,  $R_{68}$ grows thereafter due to the shrinking of the region interior to $R_{\rm cliff}$ and hence the reduction of the relative flux contribution from within small radii. At high $\alpha$ the re-expansion of $R_{68}$ is a result of viscous expansion of the low opacity region outside $R_{\rm cliff}$.

This behaviour is modified at late times when photoevaporation is included. For low $\alpha$, the evolution is qualitatively similar except that dust loss to the wind at early times results in more rapid depletion of solids by radial drift outside $R_{\rm cliff}$, and smaller maximum grain sizes, as illustrated in  Figures \ref{fig:size_suppression} and \ref{fig:f_wind_dual}. This means that the region interior to $R_{\rm cliff}$ continues to dominate the flux down to lower $R_{\rm cliff}$ values and hence sources attain a smaller minimum radius before re-expansion.
Fig. \ref{fig:flux-rad} shows that the radius from which $R_{68}$ starts to re-expand depends on $G_0$ but is relatively insensitive to the disc mass or initial radius. In the context of such low $\alpha$ models, a compact disc ($R_{68} < 20$ au) would imply either a very compact initial configuration (initial $R_{\rm C} < 30$ au) or else a wide range of initial disc sizes and radii combined with an ultraviolet field $> 100~{\rm G_0}$.
For high $\alpha$ models, the smaller grains mean that more dust is retained beyond $R_{\rm cliff}$ and hence the flux outside $R_{\rm cliff}$ dominates at late times. There is however no viscous expansion of $R_{68}$ at this   point since the outer edge of the disc is set by the wind base. At this stage the disc flux slowly declines as a result of viscous accretion of small dust on to the star but $R_{68}$ remains approximately constant. For high $\alpha$ models as well, therefore, a long lived population of small discs ($R_{68} < 20$ au) is a hallmark of the importance of photoevaporation.
 
We also plot in Fig. \ref{fig:flux-rad} the observational data of \citet{Eisner18} for the Orion Nebula Cluster (using flux values from which the estimated contributions from free-free emission have been subtracted as in \citet{Eisner18} and further computing equivalent flux values for a nominal distance of $140$ pc). In contrast to the regions observed by \citet{Tripathi}, \citet{Andrews_2018} (whose data is represented by the blue band in Fig. \ref{fig:flux-rad}), the Orion Nebula Cluster is a highly irradiated region with estimated flux levels ranging from $>10^5~{\rm G_0}$ to $< 1000~{\rm G_0}$ (though with some uncertainty based on the level of internal extinction in the cluster).
   
We see that the observational data from the Orion Nebula Cluster is broadly consistent with the low $\alpha$ models. At $G_0=10^4$, these models are brighter at a given disc size compared with non-irradiated models because photoevaporation drives more rapid disc shrinkage and there is less time for the residual disc to drain its dust by radial drift
\footnote{Note that for the high $\alpha$ models, where radial drift is less efficient, the predicted submm fluxes at given disc size for $G_0 = 10^4$ are even higher  and in excess of those observed}.
The observed data also includes objects that lie at larger sizes and lower flux values than would be predicted by our high $G_0$ models, though these may be attributable to objects that, due to a mixture of extinction and projection effects, experience a lower FUV flux.
    
Finally, we note (from the absence of black 1 Myr markers on the models with $G_0 = 10^4$ that the disc shrinkage is extremely rapid at these flux levels. This is a re-statement of the well-known `proplyd lifetime problem' in Orion, whose solution is generally held to imply that observed discs in Orion have been exposed to such strong ultraviolet fields over a small fraction of the cluster lifetime (see e.g. \citet{Winter_2019b} and references therein). 
     
Consequently, we conclude that in order for a range of FUV fluxes to reproduce the spread in the observational data from the Orion Nebula Cluster, the discs subject to a high $G_0$ must be very young, whereas those in milder environments should be older than $1$ Myr. This is consistent with the arguments of \citet{Winter_2019b} that for sufficiently low star formation efficiencies, dynamical encounters lead to preferentially younger systems in higher $G_0$ environments, and that interstellar extinction early in the cluster's evolution could have ensured that older systems have stayed in lower $G_0$ environments throughout their evolution.

\section{Discussion: implications for planet formation}
\label{sec:planet_form}
Rocky planet formation is likely to require enhancement of the mid-plane dust to gas ratio to of order of a few tens of per cent in order to be able to trigger the streaming instability and consequent gravitational fragmentation of the dust layer. Several authors have suggested that photoevaporation may provide a mechanism for  differential gas/dust removal, leaving behind a disc with suitably enhanced solid to gas ratio, conducive to planet formation \citep{Throop,Carrera_2017}. The fact that our simulations do {\it not} demonstrate significantly enhanced dust to gas ratio in the inner disc can be readily understood by inspection of the second panel of the leftmost column of Fig. \ref{fig:dust_evolution_comparison}. The possible enhancement of the dust to gas ratio only occurs at the point when the dust has grown to a size scale where it is inefficiently removed in the wind (i.e. at the first vertical dotted line at time $6 \times 10^4$ yrs). This point however coincides with an acceleration of the dust flow on to the star by radial drift, so that within a further $10^5$ years the disc is very dust depleted (see Fig. \ref{fig:d2g_vsCarrera}). This is a generic property of photoevaporation models: dust that has grown to the point that it cannot be entrained in the wind is also dust that drifts rapidly on to the star.
      
This conclusion does not preclude the possibility of achieving high dust to gas ratios if there is some mechanism for preventing the radial inflow of dust \citep{Pinilla_2012}. One such possibility is associated with the pile up of grains at the water snow line in protoplanetary discs, where the lowered fragmentation velocity for ice-free grains lowers the maximum grain size and hence inhibits radial drift within the snow line \citep{Drazkowska_2016,Schoonenberg_2017}.  While we do not follow the detailed chemodynamics of ice mantle desorption at the water snow line (typically around $2$ au for solar type pre-main sequence stars) we can use the results of \citet{Ormel_2017} which examines the maximum dust to gas ratio achievable at the snow line as a function of the ratio of dust to gas fluxes arriving from the outer disc. This study found that a minimum ratio of order unity is required in order to achieve suitable pre-conditions for the streaming instability.
      
\begin{figure}
    \centering
        \includegraphics[width=\linewidth]{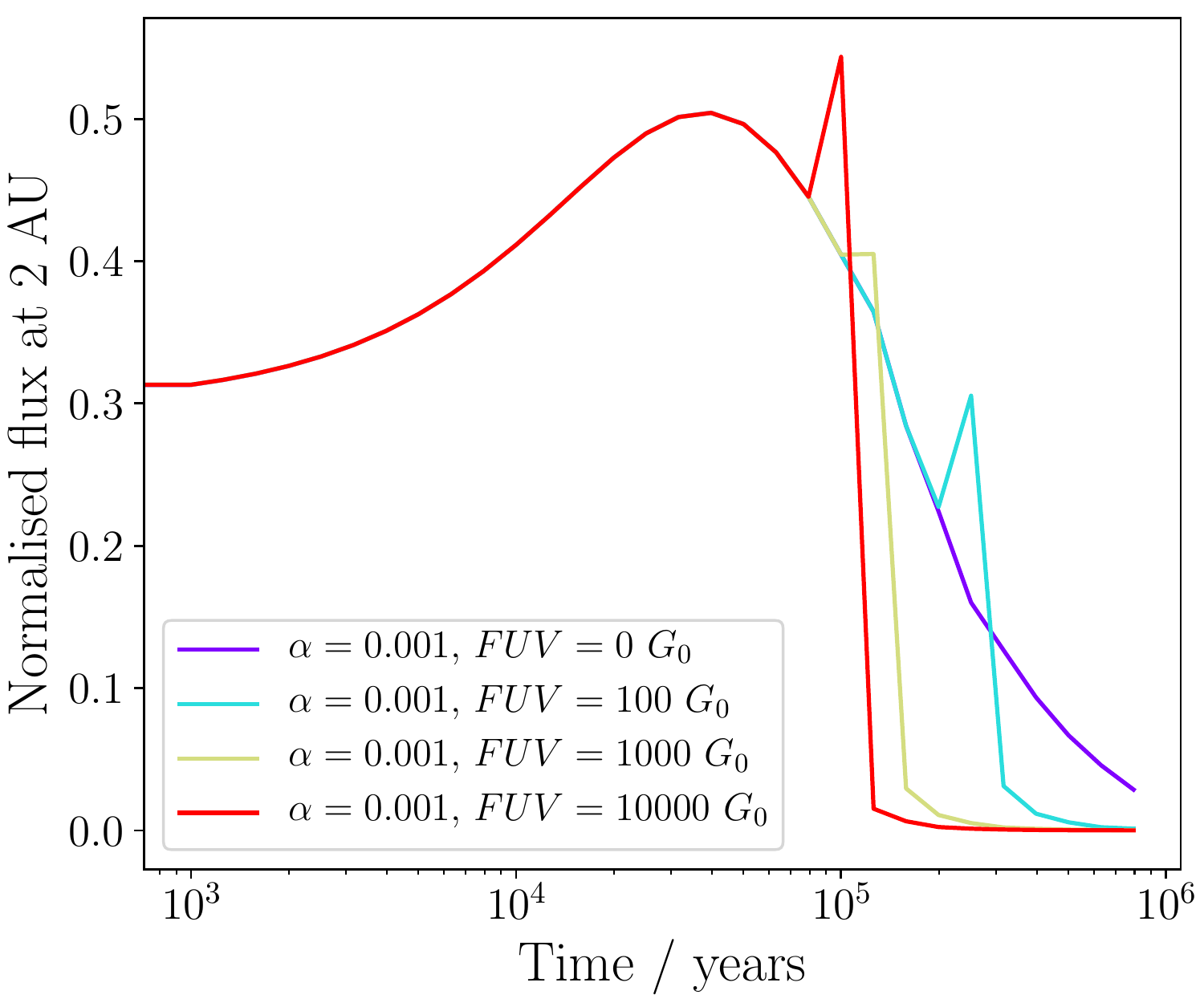}
        \caption{The flux of drifting dust normalised by the viscous accretion rate of the gas  as a function of age. FUV flux increases from the purple (no FUV) to the red ($10^4~{\rm G_0}$).}
        \label{fig:normalised_flux}
\end{figure}

In Fig. \ref{fig:normalised_flux} we show a suite of models with $\alpha = 10^{-3}$ {\footnote{Note that for $\alpha=10^{-2}$ grains and gas remain sufficiently well coupled that there is no significant enhancement of the normalised flux.}} for different FUV flux levels and see that in the absence of photoevaporation this condition can be satisfied at early times thanks to the short growth and drift time-scales in the inner disc. After around $10^5$ years the depletion of disc dust due to radial drift results in a decline of this flux ratio; in the absence of photoevaporation, the dust to gas flux ratio falls below $10$ per cent at an age of $\sim 3 \times 10^5$ years. Fig. \ref{fig:normalised_flux} shows that photoevaporation causes a steeper decline in the dust to gas flux ratio, an effect that can be attributed to the previous removal of dust from the outer disc which reduces the available reservoir flowing in to the inner disc. This narrows the window during which it is likely that the streaming instability can be triggered at the water snow line but the dependence on FUV field is weak (e.g. an increases of FUV background from $100~{\rm G_0}$ to $10^4~{\rm G_0}$ reduces the time at which the flux ratio drops below $0.1$ by only a factor three.
      
We thus conclude that the potential to form planets at the water snow line by this mechanism is not severely affected by even high levels of photoevaporation, provided that this process occurs within $\sim 10^5$ years. On the other hand, photoevaporation is clearly a mild inhibitor of planet formation  both because of the solid mass lost in the wind  (see Fig. \ref{fig:f_wind_dual}) and because of the shortening of the epoch of enhanced normalised flux at the snow-line (Fig. \ref{fig:normalised_flux}). See \citet{Haworth_18_Trappist} for a discussion of how photoevaporation would place very stringent requirements on planet formation efficiency in the Trappist 1 system where the mass in rocky planets is a significant fraction of the initial disc mass contained in solids.
      
\section{Conclusions}
Our study of dust dynamics in protoplanetary discs subject to photoevaporation by  external FUV radiation has the following principal conclusions:

   i) Dust loss in the wind is limited by the time required
for the maximum grain size in the outer disc to grow to a size
where it cannot be entrained due to strong radial drift. Typically dust stops being entrained within $\sim 10^5$ years at which point it has grown to $10-100~\mu$m. For higher $\alpha$ models, the dust may not grow large enough that this condition is met (Fig. \ref{fig:Stokes_optimum}).

  ii) From this point onwards, dust in the outer disc is depleted
by inward radial drift (Fig. \ref{fig:dust_evolution_comparison}). This is  particularly  the case in  discs with
low turbulence levels ($\alpha = 10^{-3}$)  where grains can grow to larger sizes.  Higher levels of turbulence  ($\alpha = 10^{-2}$) keep grains smaller and hence more tightly coupled to the gas and so dust depletion is less severe. 

 iii) The rate of dust depletion by radial drift is strongly
dependent on
the strength of the external FUV field since this determines how much
of the dust reservoir at large radius has been lost to the wind. Early
loss of outer disc grains  in the wind prevents the  
re-supply of dust to the inner
disc which would otherwise partially offset  the effect of radial drift. Thus
although the fraction of disc dust lost to the wind never exceeds 
$\sim 60$ per cent in our models, this can have profound effects on the
dust depletion  time-scale in the disc.
             
 iv)  Disc demographic studies indicate that the  time-scale for
dust depletion (as measured by  $2~\mu$m excess
emission) is not significantly shorter than the time over which
accretion on to the star declines. This requirement is only met by  our
models for discs in which the Shakura Sunyaev $\alpha$ 
turbulence parameter is relatively high ($10^{-2}$) since the smaller
grain sizes in this case result in closer coupling between the dust and
gas evolution (Fig. \ref{dustgasl}). Although both dust and accretion lifetimes decline with
increasing ultraviolet field strengths, their {\it relative} values
depend on $\alpha$ rather than the strength of the external FUV field.

 v) Conversely the predicted trajectories of models in the plane of disc radius versus  flux (both measured at $850~\mu$m) are better matched to observations at all FUV levels if  $\alpha$ is low ($\sim 10^{-3}$).
 In particular the models with  high FUV levels pass through the region of parameter space (with relatively high flux to radius ratios) occupied by discs observed in the highly irradiated environment of the Orion Nebula Cluster.

vi) The predominant factor in setting the location of the base of the photoevaporative wind is the strength of the external FUV field. Conversely, factors such as the initial disc mass and radius and viscosity have little effect. The trajectories in the plane of dust disc radius and submm flux from the dust for a given $\alpha$ and FUV environment become, in photoevaporating environments, likewise degenerate across a range of initial conditions (Fig. \ref{fig:flux-rad}). 

 vii) Since the consequence of photoevaporation is to {\it decrease} the lifetime of
dust in discs, it is over all a negative factor with respect to forming rocky planets.
However it has only a very mild impact on models in which rocky planets form from
drifting solids at the water snow-line (Fig. \ref{fig:normalised_flux}); such models predict a peak flux of solids
at an early stage of disc evolution ($\ll$ a Myr) when the impact of photoevaporation is minor.

 viii) Photoevaporation does {\it not} result in an enhancement of the solid to gas ratio
in the outer disc (Fig. \ref{fig:d2g_vsCarrera}). Dust that has grown large enough not to be entrained in the wind
is instead subject to efficient radial drift and the dust to gas ratio  never
significantly exceeds its initial value.

\section*{Acknowledgements}
We are grateful to Tom Haworth, Giovanni Rosotti and Andrew Winter for useful discussions.
We thank the anonymous reviewer for their feedback which helped us clarify our modelling assumptions.
RB and CJC acknowledge support from the STFC consolidated grant ST/S000623/1.
This work has also been supported by the European Union's Horizon 2020 research and innovation programme under the Marie Sklodowska-Curie grant agreement No 823823 (DUSTBUSTERS).
\addcontentsline{toc}{section}{Acknowledgements}




\bibliographystyle{mnras}
\bibliography{bib} 




\appendix
\section{Comparison of Mass Loss Implementations}

\begin{figure*}
    \centering
    \includegraphics[width=\linewidth]{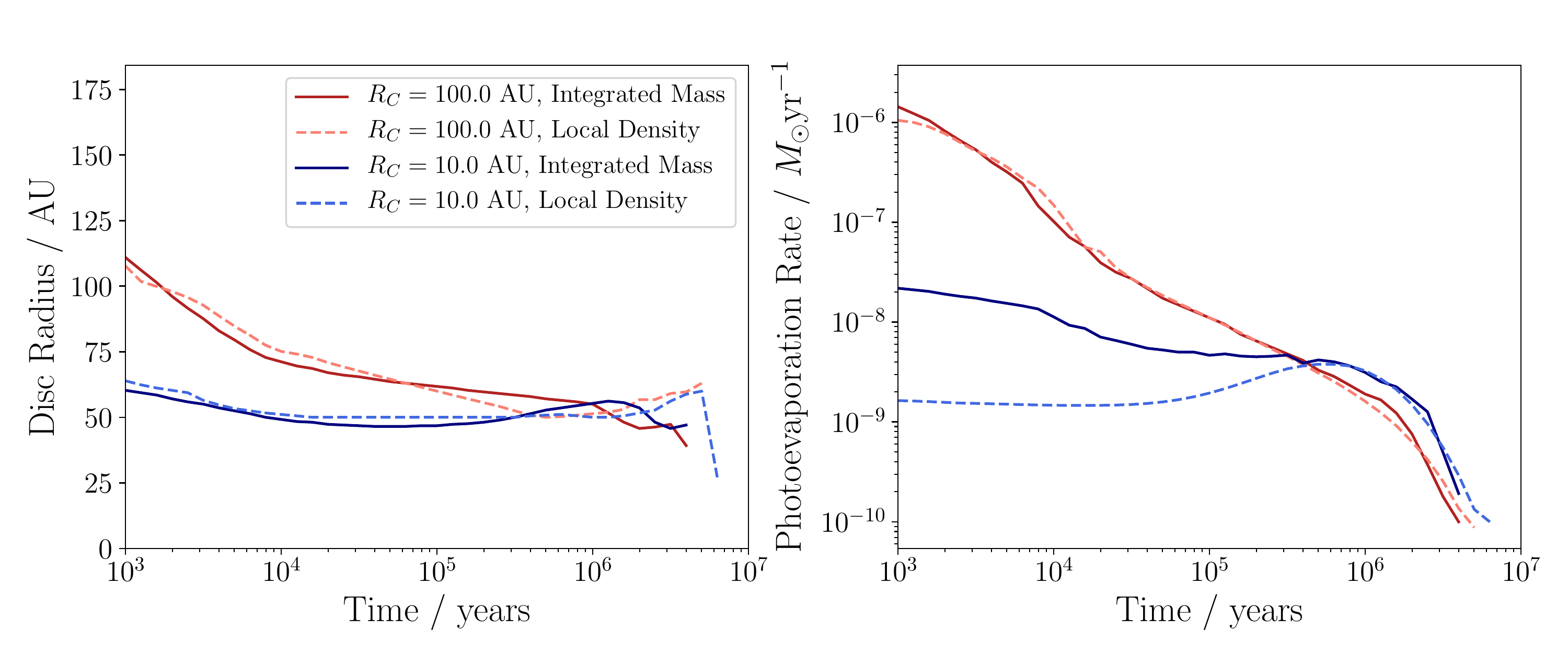}
    \caption{Left: The evolution of disc radius measured using the position of $\dot{M}_{\rm max}$. Right: the evolution of the photoevaporative mass loss rate. Solid lines represent discs with photoevaporation calculated using the total mass, whereas dashed lines represent those models using the local surface density at the disc edge. The red-pink colours represent models with initial scale radius $R_{\rm C} = 100$ au and blue colours $R_{\rm C} = 10$ au.}
    \label{fig:appendix_fig}
\end{figure*}
In this paper we devise a method of locating the disc outer edge and corresponding mass loss rate  in terms of the location where the wind becomes optically thick to FUV radiation (see Section 2.2.2). In this way our mass loss rates are never calculated in the (unphysical)  optically thin wind regime where there is the problem that the wind mass loss rate scales linearly with the 
surface density. In the latter case, the situation can arise where depletion of the outermost
disc grid cell is accompanied by a turning off
of the mass loss rate from the disc, ignoring
the much greater mass loss that would result if
an interior cell was deemed to be the outer edge. 

In order to avoid this problem, previous implementations of photoevaporation from the FRIED grid \citep{Winter_2019b} did not use the surface density in the
outermost grid cell but instead a notional
surface density calculated from the total
disc mass and radius assuming a power law
surface density profile; for the viscosity law assumed here
(where $\nu \propto R$) this is such that $\Sigma_{\rm out}$ satisfies $M_{\rm D} = 2\pi \Sigma_{\rm out} R_{\rm out}^2$.

The benefit of using this method is that the photoevaporation rates depend on a global quantity, and are insensitive to numerical issues at the disc edge. However, this also poses a risk of calculating the wrong mass loss rate if the relationship between the local conditions at the disc edge and the total mass is different from that assumed (for example in cases where the density profile develops an exponential tail as in the viscous similarity solution of  \citet{LBP_1974}).

Here we check whether the previous implementation of \citep{Winter_2019b} produces significantly different results from our updated prescription for `typical' disc evolutionary scenarios. To this end we compared the prescriptions using  models of gas-only discs, one with scale radius $R_{\rm C} = 100$ au and the other with $R_{\rm C} = 10$ au, both with initial mass $M_{\rm D} = 0.02~M_{\sun}$ and subjected to an FUV flux of 1000 ${\rm G_0}$.

The left panel of Fig. \ref{fig:appendix_fig} illustrates how the behaviour of the radius is qualitatively similar between the two methods, regardless of initial disc size, showing familiar shrinking, stalling and spreading phases.
For initially large discs, the mass loss rates were in excellent agreement throughout the lifetime of the model.
For initially compact discs, there was a discrepancy by over an order of magnitude in the mass loss rates at early times although at later times $\gtrsim$ 1 Myr, the mass loss rates did converge between the methods.

The more compact discs start with a wind base that lies well outside the exponential cut off radius in the initial density profile, meaning that the power law conversion between density and mass used by \citet{Haworth_18_FRIED} is not appropriate. In initially extended discs, or at late times after significant viscous spreading, the steady state profile holds throughout the disc, so the different methods agree.
We conclude that the method used by \cite{Winter_2019b}  works well for large discs but can run into difficulties if, for whatever reason, the disc surface density profile deviates from the assumed power law. While in many cases the differences are not significant, the method employed in this paper is to be preferred on account of its less restrictive assumptions.



\bsp	
\label{lastpage}
\end{document}